\documentclass[11pt]{article}
\usepackage{amsmath,amssymb,graphicx}
\usepackage{hyperref}
\usepackage[nosort]{cite}
\usepackage{multicol,color,longtable}
\usepackage{arydshln}

\usepackage{cite}
\usepackage{amsmath,amsfonts,amssymb,calrsfs}
\usepackage[small,bf,hang]{caption}
\usepackage{slashed}
\usepackage{latexsym,epsfig}

\definecolor{darkred}{rgb}{0.65,0.15,0}
\definecolor{darkgreen}{rgb}{.05,.5,.05}
\hypersetup{pdfborder={0 0 0},colorlinks=true,urlcolor=darkred,citecolor=blue,linkcolor=darkred,linktocpage=true}

\usepackage[final]{showkeys}
\usepackage{amsfonts}
\usepackage{mathrsfs}
\usepackage{dsfont}
\usepackage[Symbol]{upgreek}

\DeclareMathAlphabet{\mathpzc}{OT1}{pzc}{m}{it}

\newcommand{\be}{\begin{align}}
\newcommand{\ee}{\end{align}}

\makeatletter

\@addtoreset{equation}{section}
\makeatother

\newcommand{\bea}{\begin{eqnarray}}
\newcommand{\eea}{\end{eqnarray}}

\def\ss{\scriptstyle}

\def\*{\partial}

\def\={\!=\!}

\usepackage{cite}
\usepackage{amsmath,amsfonts,amssymb}%,calrsfs}

\usepackage{color}
\definecolor{red}{rgb}{1,0,0}
\definecolor{lred}{rgb}{0.3,0,0}
\definecolor{green}{rgb}{0,0.6,0}
\definecolor{blue}{rgb}{0,0,1}
\definecolor{violet}{rgb}{0.8,0,0.8}

\newcommand{\Ref}[1]{(\ref{#1})}

\begin{document}

\begin{titlepage}

\vfill
\begin{flushright}
%PREPR-IN-T
\end{flushright}

%\vfill

\begin{center}

\mathversion{bold}

   \baselineskip=16pt
   {\LARGE \bf  \mbox{Consistent $S^3$ reductions of six-dimensional supergravity}}
   \vskip 2cm

\mathversion{normal}

{\large\bf {Henning Samtleben${\,}^1$, \"{O}zg\"{u}r Sar{\i}o\u{g}lu${\,}^2${}}}
\vskip .6cm
{\it ${}^1$
Univ Lyon, Ens de Lyon, Univ Claude Bernard, CNRS,\\
Laboratoire de Physique, F-69342 Lyon, France} \\
{henning.samtleben@ens-lyon.fr}
\vskip .2cm

{\it ${}^2$
Department of Physics,
Middle East Technical University, \\
06800, Ankara, Turkey}\\
sarioglu@metu.edu.tr
\vskip .2cm

\end{center}

%\vfill 

\vspace*{1cm}

\begin{center} 
\textbf{Abstract}

\end{center} 
\begin{quote}

We work out the consistent AdS$_3\times S^3$ truncations of the bosonic sectors of
both the six-dimensional ${\cal N}=(1,1)$ and ${\cal N}=(2,0)$ supergravity theories.
%respectively, to three dimensions.
They result in inequivalent three-dimensional half-maximal ${\rm SO}(4)$ 
gauged supergravities describing 32 propagating bosonic degrees of freedom apart from
the non-propagating supergravity multiplet. We present the full
non-linear Kaluza-Klein reduction formulas and illustrate them by explicitly uplifting a number of AdS$_3$ vacua.

\end{quote} 
\vfill
\setcounter{footnote}{0}
\end{titlepage}

\clearpage
\setcounter{page}{2}

\tableofcontents

%%%%%%%%%%%%%%%%%%%%%%%%%%%%%%%

\section{Introduction}

%%%%%%%%%%%%%%%%%%%%%%%%%%%%%%%

Consistent sphere truncations have a long history in supergravity.
Within maximal supergravity, this goes back to the seminal 
work of \cite{deWit:1986iy} on the consistent truncation of eleven-dimensional
supergravity on AdS$_4\times S^7$ to the lowest Kaluza-Klein multiplet 
giving rise to four-dimensional ${\rm SO}(8)$ gauged supergravity.
An analogous result for AdS$_7\times S^4$ was established in \cite{Nastase:1999kf},
while the proof of the consistent truncation of IIB supergravity on AdS$_5\times S^5$
was completed only recently~\cite{Baguet:2015sma}.
Consistent truncations have led to a better comprehension of the structures of
the theories of concern and the dualities they enjoy. Notably, these are not truncations 
in an effective field theory sense, with the massive Kaluza-Klein towers integrated 
out, yet every solution of the lower-dimensional theory lifts to a solution of the 
higher-dimensional theory. They are of particular importance in holographic 
applications, ensuring the validity of lower-dimensional supergravity computations,
such as holographic correlators and renormalization group (RG) flows~\cite{Bianchi:2001de}.

This work deals with consistent sphere compactifications in the context of AdS$_3\times S^3$,
one of the central examples in the AdS/CFT correspondence~\cite{Maldacena:1997re}
in which supergravity techniques have been successfully employed
\cite{deBoer:1998kjm,Deger:1998nm,Mihailescu:1999cj,Arutyunov:2000by,Giusto:2018ovt,Rastelli:2019gtj}
in order to unravel the structure of the dual two-dimensional conformal field theories.
Generic consistent $S^3$ truncations in (super)gravity have been discussed in 
\cite{Cvetic:2000dm, Cvetic:2000zu,Cvetic:2000ah},
where the full non-linear Kaluza-Klein Ans\"atze were constructed for a higher-dimensional
theory that comprises the field content of the bosonic string.
The resulting lower-dimensional theories are ${\rm SO}(4)$ gauged (super)gravities carrying 
gauge fields, a 2-form, and scalar fields whose potential do not admit any stationary points.
In the particular case of AdS$_3\times S^3$, the higher-dimensional theory is 
$D=6$, ${\cal N}=(1,0)$ supergravity coupled to a single tensor multiplet that carries an 
anti-selfdual 2-form. In contrast to the higher-dimensional examples, the 2-forms in the resulting 
three-dimensional theory are auxiliary and can be integrated out, giving rise to an 
additional contribution to the scalar potential. This amended potential turns out to support a 
stable supersymmetric AdS$_3$ vacuum \cite{Deger:2014ofa}, corresponding to the supersymmetric 
${\rm AdS}_3 \times S^3$ solution of the $D=6$ theory.
The non-linear Kaluza-Klein Ans\"atze can be confirmed by direct computation.

More recently, new techniques have emerged for a more systematic understanding of consistent
truncations within exceptional field theory (ExFT) and generalized 
geometry~\cite{Hohm:2013pua,Lee:2014mla,Hohm:2014qga,Malek:2017njj,Inverso:2017lrz,Cassani:2019vcl}, 
see also \cite{Anderson:2017uxt,Hassler:2019wvn} in the context of double field theory.
Using the reformulation of $D=6$, ${\cal N}=(1,0)$ supergravity as an exceptional field theory 
based on the group ${\rm SO}(4,4)$ \cite{Hohm:2017wtr}, the non-linear Kaluza-Klein 
Ans\"atze from \cite{Cvetic:2000dm,Deger:2014ofa} can straightforwardly be reproduced 
from the generalized Scherk-Schwarz twist matrices ${\cal U}$ in this framework.
In this paper, we will extend the consistent $S^3$ truncations to the full ${\cal N}=(1,1)$ 
and ${\cal N}=(2,0)$ supergravities in six dimensions.
The relevant framework is an ${\rm SO}(8,4)$ ExFT which, 
depending on the solution of its section constraint, 
describes the aforementioned six-dimensional supergravities \cite{Hohm:2017wtr}.
The resulting three-dimensional theories are ${\rm SO}(4)$ gauged supergravities coupled to 
4 half-maximal scalar multiplets~\cite{Nicolai:2001ac,deWit:2003ja}, i.e.\
with scalar target space given by ${\rm SO}(8,4)/\big({\rm SO}(8)\times {\rm SO}(4)\big)$.
In the ExFT framework, the construction of the consistent truncations simply 
amounts to embedding the ${\rm SO}(4,4)$ twist matrices ${\cal U}$
into the ${\rm SO}(8,4)$ isometry group of the ungauged three-dimensional theory.
Two inequivalent embeddings give rise to inequivalent three-dimensional 
gaugings, describing the truncation of the ${\cal N}=(1,1)$ and the ${\cal N}=(2,0)$ theory, respectively.
\smallskip

The paper is organized as follows. In section~\ref{sec:3D}, we introduce the relevant 
three-dimensional supergravities, their gauge structure and scalar potentials. We give an 
explicit parametrization of their scalar target space 
${\rm SO}(8,4)/\big({\rm SO}(8)\times {\rm SO}(4)\big)$ and determine the full set of 
stationary points of their scalar potentials.
In section \ref{sec:ExFT}, we review the framework of ${\rm SO}(8,4)$ ExFT. 
In particular, we discuss the two inequivalent solutions of its section constraint and establish 
the full dictionary of the ExFT fields into the six-dimensional fields of 
${\cal N}=(1,1)$ and ${\cal N}=(2,0)$ supergravity, respectively.
In sections~\ref{sec:N11uplift} and~\ref{sec:N20uplift}, we use the explicit 
Scherk-Schwarz twist matrix ${\cal U}$ together with the ExFT-supergravity dictionary 
to work out the full non-linear Kaluza-Klein Ans\"atze for all six-dimensional fields, defining 
the consistent truncation. As an illustration and a consistency check, we use these 
Ans\"atze in section~\ref{sec:uplifts} in order to give the explicit uplift of some of the
three-dimensional AdS$_3$ vacua into full solutions of $D=6$ supergravity.
We close with some comments in section~\ref{sec:conclusions}.

%%%%%%%%%%%%%%%%%%%%%%%%%%%%%%%

\section{The three-dimensional supergravity}
\label{sec:3D}

%%%%%%%%%%%%%%%%%%%%%%%%%%%%%%%

In this section, we collect the basic formulas of the relevant three-dimensional
supergravities. In particular, we give an explicit parametrization of their scalar target space,
which allows us to determine the full set of stationary points of the scalar potentials.

\subsection{3D gauged supergravity}
\label{subsec:3D}

Three-dimensional gauged supergravity with ${\cal N}=8$ (half-maximal) supersymmetry 
has been constructed in \cite{Nicolai:2001ac,deWit:2003ja}.
The theory is based on the coset space
\bea
{\rm G}/{\rm H} &=& {\rm SO}(8,4)/\big({\rm SO}(8)\times {\rm SO}(4)\big)
\;,
\label{coset}
\eea
with all couplings
completely specified by the choice of a constant symmetric embedding tensor $\Theta_{\bar K\bar L,\bar M\bar N}$
of the form
\bea
\Theta_{\bar K\bar L,\bar M\bar N}&=&\theta_{\bar K\bar L\bar M\bar N}+
\frac12\,\left(
\eta_{\bar M[\bar K} \theta_{\bar L]\bar N}-
\eta_{\bar N[\bar K} \theta_{\bar L]\bar M}
\right)
+\theta\,\eta_{\bar M[\bar K}\eta_{\bar L]\bar N}
\;,
\label{embedding0}
\eea
with antisymmetric $\theta_{\bar K\bar L\bar M\bar N}=\theta_{[\bar K\bar L\bar M\bar N]}$, symmetric $\theta_{\bar M\bar N}=\theta_{(\bar M\bar N)}$
and the ${\rm SO}(8,4)$ invariant tensor $\eta_{\bar M\bar N}$\,.
Indices ${\bar M}, {\bar N}, \dots$ label the vector representation of ${\rm G}={\rm SO}(8,4)$,
and are raised and lowered with $\eta_{\bar M\bar N}$\,.
The embedding tensor encodes the minimal coupling of vector fields to scalars according to
\bea 
D_\mu M_{\bar M\bar N} &\equiv& \partial_\mu M_{\bar M\bar N} +
2\,A_\mu{}^{\bar P\bar Q}\,\Theta_{\bar P\bar Q,\bar K\bar L}\,(T^{\bar K\bar L})_{(\bar M}{}^{\bar R}  M_{\bar N)\bar R} 
\;, \label{DS} 
\eea 
with the symmetric matrix $M_{\bar M\bar N}$ parametrizing the coset space (\ref{coset}).
By $T^{\bar M\bar N}$ we denote the generators of
${\mathfrak g}={\rm Lie}\,{\rm G}$ acting by left multiplication with the algebra
\bea
{}[T^{\bar K\bar L},T^{\bar M\bar N}] &=&
2\,(\eta^{\bar K[\bar M}\,T^{\bar N]\bar L}-\eta^{\bar L[\bar M}\,T^{\bar N]\bar K}) \;.
\label{SO}
\eea
The number of vector fields involved in the connection \Ref{DS}
is equal to the rank of $\Theta_{\bar K\bar L,\bar M\bar N}$
(taken as a ${\rm dim\,G}\times {\rm dim\,G}$ matrix).

The complete bosonic Lagrangian of the
three-dimensional theory is given as a gravity coupled Chern-Simons gauged 
${\rm G}/{\rm H}$ coset space $\sigma$-model
\bea
e^{-1}{\cal L} &=& \frac1{4} R
+ \frac1{32} \,g^{\mu\nu}\,\partial_\mu M^{\bar M\bar N} \partial_\nu M_{\bar M\bar N}
+ e^{-1}{\cal L}_{\rm CS} - V \;,
\label{LCS}
\eea
with three-dimensional metric $g_{\mu\nu}$ and $e\equiv\sqrt{|{\rm det}\,g_{\mu\nu}|}$\,.
The Chern-Simons term is explicitly given by
\bea
{\cal L}_{\rm CS}&=&
\frac14\varepsilon^{\mu\nu\rho} A_\mu{}^{\bar K\bar L} \,\Theta_{\bar K\bar L,\bar M\bar N}\,
\left( \partial_\nu A_\rho{}^{\bar M\bar N}
+ \frac13\,
f^{\bar M\bar N,\bar P\bar Q}{}_{\bar R\bar S}\,\Theta_{\bar P\bar Q,\bar U\bar V}\,A_\mu{}^{\bar R\bar S} A_\rho{}^{\bar U\bar V}\right)
\;,
\eea
in terms of the embedding tensor (\ref{embedding0}),
with the ${\rm SO}(8,4)$
structure constants $f^{\bar M\bar N,\bar P\bar Q}{}_{\bar R\bar S}$ from \Ref{SO}. The form of the scalar potential $V$ is 
determined by the embedding tensor and may be written in the form \cite{Schon:2006kz}
(where we have corrected a typo in the second line)
\bea
   V &=&   \frac1{48} \,  \theta_{\bar K\bar L\bar M\bar N} \theta_{\bar P\bar Q\bar R\bar S} \Big(
              M^{\bar K\bar P} M^{\bar L\bar Q} M^{\bar M\bar R} M^{\bar N\bar S}
	           -6\, M^{\bar K\bar P} M^{\bar L\bar Q} \eta^{\bar M\bar R} \eta^{\bar N\bar S}
	           \nonumber \\
	           &&{}\qquad\qquad\qquad\qquad\qquad
		   +8 \, M^{\bar K\bar P} \eta^{\bar L\bar Q} \eta^{\bar M\bar R} \eta^{\bar N\bar S}
		   -3 \, \eta^{\bar K\bar P} \eta^{\bar L\bar Q} \eta^{\bar M\bar R} \eta^{\bar N\bar S}
	     \Big) \nonumber \\
	           &&{}
	          + \frac1{32} \, \theta_{\bar K\bar L} \theta_{\bar P\bar Q} \Big(
	          2 M^{\bar K\bar P} M^{\bar L\bar Q} - 2 \eta^{\bar K\bar P} \eta^{\bar L\bar Q}
	          -M^{\bar K\bar L} M^{\bar P\bar Q} \Big) 
	        %  \nonumber \\
	         % &&{} 
	          + \theta \, \theta_{\bar K\bar L} M^{\bar K\bar L} - 8\, \theta^2  
    \label{NVD3}	   
    \;.
\eea
From the general expression of the scalar potential, we have omitted the term 
carrying a totally antisymmetric $M^{\bar K\bar L\bar M\bar N\bar P\bar Q\bar R\bar S}$
which drops out upon restriction to embedding tensors 
satisfying the additional constraint
\bea
\Theta_{[\bar K\bar L,\bar M\bar N} \,\Theta_{\bar P\bar Q,\bar R\bar S]}&=& 0
\;.
\label{extraQ}
\eea
As pointed out in \cite{Hohm:2017wtr}, consistent truncations obtained by generalized Scherk-Schwarz reduction
necessarily lead to three-dimensional theories satisfying (\ref{extraQ}),
and we will in the following restrict to such theories.
For the fermionic completion of (\ref{LCS}) and its 
full supersymmetry transformations we refer to \cite{Nicolai:2001ac,deWit:2003ja}.

For the following, it will be convenient to choose a specific basis upon breaking
\bea
{\rm SO}(8,4) &\longrightarrow& 
{\rm GL}(4) \times {\rm SO}(4)
\;,
\nonumber\\
X^{\bar M} &\longrightarrow& \left\{
X^A, X_A, X^\alpha \right\}
\;,
\label{GL4SO4}
\eea
with $A=1, \dots, 4$ and $\alpha=1, \dots, 4$ labelling the ${\rm GL}(4)$ and the ${\rm SO}(4)$
vector representation, respectively. In this basis, the ${\rm GL}(4)$ is embedded into an ${\rm SO}(4,4)$, such that the ${\rm SO}(8,4)$ invariant tensor is of the form
\bea
\eta_{\bar M\bar N}&=&
\begin{pmatrix}
0 & \delta_A{}^B&0 \\
 \delta_B{}^A &0 &0\\
 0&0& - \delta_{\alpha\beta}
 \end{pmatrix} 
 \;.
 \label{MAB}
 \eea
Specifically, we will be interested in the theories described by the two embedding tensors
\bea
\mbox{(A)} &:&
\theta_{AB} = 4 \, \delta_{AB} \;, \quad
\theta_{ABCD} = -2\,\alpha\,\varepsilon_{ABCD}
\;,
\nonumber\\
\mbox{(B)} &:&
\theta_{ABC}{}^{D} = \varepsilon_{ABCE}\,\delta^{ED}\;,\quad
\theta_{ABCD} = -2\,\alpha\,\varepsilon_{ABCD}
\;,
\label{embeddingAB}
\eea
with the totally antisymmetric $\varepsilon_{ABCE}$,
a free constant $\alpha$. These theories capture the $S^3$ reductions of ${\cal N}=(1,1)$ and ${\cal N}=(2,0)$
supergravity, respectively.
In particular, the embedding tensors induce the gauge connections
\bea
-\frac14\,A_\mu{}^{\bar K\bar L}\,\Theta_{\bar K\bar L,\bar M\bar N}\,T^{\bar M\bar N} &=&
\left\{
\begin{array}{lr}
A_{\mu}{}^{AB}\,   T_B{}^{A}
+\left(
A_{\mu\,B}{}^{A}+\frac{\alpha}{2}\, \varepsilon_{ABCD}\,A_\mu{}^{CD} \right) T^{AB}
&
\mbox{(A)}\\[2ex]
\tilde A_\mu{}^{AB} \,T_B{}^A 
+\left(
A_{\mu\,B}{}^A+
\frac{\alpha}{2}\, \varepsilon_{ABCD}  \,\tilde A_\mu{}^{CD}
 \right) \tilde T^{AB}
&\mbox{(B)}
\end{array}
\right.
\;,\qquad
\label{connectionAB}
\eea
with 
\bea
\tilde A_\mu{}^{AB}&=& \frac12\,\varepsilon^{ABCD}\,A_\mu{}^{CD}\;,\nonumber\\
\tilde T{}^{AB}&=& \frac12\,\varepsilon^{ABCD}\,T{}^{CD}\;,
\label{Atilde}
\eea
in the second case.
Both embedding tensors induce a gauge group of non-semisimple type 
\bea
{\rm G}_{\rm gauge} &=&
{\rm SO}(4)\ltimes T^6
\;,
\label{gaugeST}
\eea
with the abelian generators $\left\{T^{AB}\right\}$ of $T^6$ transforming in the adjoint representation of ${\rm SO}(4)$\,.
Chern-Simons gauge theories with gauge group of type (\ref{gaugeST}) and the $T^6$ generators realized 
as shift symmetries on scalar fields can be rewritten as ${\rm SO}(4)$ Yang-Mills theories upon integrating
out the vectors $A_{\mu\,A}{}^B$ associated to the $T^6$ generators~\cite{Nicolai:2003bp,deWit:2003ja}.

\subsection{Parametrization of the ${\rm SO}(8,4)/\big({\rm SO}(8)\times {\rm SO}(4)\big)$ scalar coset}

In order to study the structure of the scalar potential (\ref{NVD3}), it turns out to be useful to adopt particular
parametrizations of the scalar matrix $M_{\bar M\bar N}$. To this end, we decompose the ${\rm SO}(8,4)$
generators according to (\ref{GL4SO4}), such that a coset element 
${\cal V}\in {\rm SO}(8,4)/\big({\rm SO}(8)\times {\rm SO}(4)$\big)
can be parametrized in triangular gauge as
\bea
{\cal V} &=& e^{\phi_{AB} T^{AB} } e^{\phi_{A\alpha} T^{A\alpha} }  {\cal V}_{{\rm GL}(4)}
\;,
\eea
with nilpotent generators $T^{AB}=T^{[AB]}$ and a ${{\rm GL}(4)}$ matrix ${\cal V}_{{\rm GL}(4)}$. Modula an ${\rm SO}(4)$ gauge
freedom, this matrix carries the 32 physical scalar degrees of freedom. In the following, we will make use of the fact that the gauge
groups we are studying include shift symmetries acting on the scalars $\phi_{AB}$, c.f.~(\ref{connectionAB}),
which we may use to adopt a gauge in which 
$\phi_{AB}\rightarrow0$. As a result, the gauge group (\ref{gaugeST}) reduces to a standard ${\rm SO}(4)$\,.

Explicitly, we choose a representation such that 
\bea
{\cal V}_ {\bar M}{}^{\bar K}\;\big|_{\phi_{AB}\rightarrow0} &=& 
\left(
\begin{array}{c:c:c}
{\cal V}_A{}^B &\frac12\,\phi_{A\gamma}{}\phi_{C\gamma} ({\cal V}{}^{-1})_B{}^C & 
\phi_{A\gamma}{}\delta^{\gamma\beta}
\\ \hdashline
0 &({\cal V}{}^{-1})_B{}^A&0
\\ \hdashline
0 & \phi_{C\alpha}({\cal V}{}^{-1})_B{}^C
&{\delta}_\alpha{}^\beta
\end{array}
\right)
\;,
\eea
in the basis (\ref{GL4SO4}), with ${\cal V}_A{}^B\equiv({\cal V}_{{\rm GL}(4)})_A{}^B$.
The symmetric positive definite matrix ${M}_{\bar M\bar N}=
\left({\cal V}{\cal V}^T\right)_ {\bar M\bar N}$ then takes the form
\bea
{M}_{\bar M\bar N}
=
{\small
\left(
\begin{array}{c:c:c}
\!\!m_{AB} +
\frac14(\phi\phi)_{AC} m^{CD} (\phi\phi)_{DB}
 + 
(\phi\phi)_{AB}
&
\frac12(\phi\phi)_{AC} m^{BC} \!\!
&
\frac12\,(\phi\phi)_{AC} m^{CD}  \phi_{D\beta}
+ 
\phi_{A\beta}\!\!\!
\\ \hdashline
\frac12(\phi\phi)_{BC} m^{AC}
&m^{AB}& 
\phi_{C\beta}m^{AC}
\\ \hdashline
\frac12\,\phi_{C\alpha} \, m^{CD} \,(\phi\phi)_{BD}
+ \phi_{B\alpha}
&
\phi_{C\alpha}m^{BC}
&
\delta_{\alpha\beta}+
\phi_{C\alpha} \phi_{D\beta}\,m^{CD}
\end{array}
\right)}
\;,
\nonumber\\
\label{paramM}
\eea
with $m_{AB}\equiv{\cal V}_A{}^C{\cal V}_B{}^C$, $m^{AB}$ denoting its inverse matrix
$m_{AC}m^{CB}=\delta_A{}^B$, and where we have denoted 
$(\phi\phi)_{AB}\equiv \phi_{A\gamma}\phi_{B\gamma}$\,.

Evaluating the scalar kinetic term from (\ref{LCS}) in this parametrization yields
\bea
{\cal L}_{\rm kin} &=&
-\frac1{32}\,
{\rm Tr}\left[
D_\mu M \,M^{-1} \,D^\mu M \,M^{-1}
\right]
\label{kinetic3D}\\
&=&
-\frac{1}{16}\,{\rm Tr}\left[
D_\mu m \,m^{-1}\, D^\mu m\, m^{-1}
\right]
-\frac{1}{8}\,\,D_\mu \phi_{A\alpha}\,m^{AB}\,D^\mu \phi_{B\alpha}
+\frac1{64}\,Y_{\mu\,AB} m^{BC} Y^\mu{}_{CD} m^{DA}
\;,
\nonumber
\eea
with 
\bea
Y_{\mu\,AB} &=& D_\mu \phi_{A\alpha} \phi_{B\alpha} - D_\mu \phi_{B\alpha} \phi_{A\alpha}
\;,
\eea
and ${\rm SO}(4)$ covariant derivatives $D_\mu$.
The first term in (\ref{kinetic3D}) represents a ${\rm GL}(4)/{\rm SO}(4)$ $\sigma$-model.

Let us finally evaluate the scalar potential for the two choices of embedding tensor
(\ref{embeddingAB}).
For the embedding tensor (A), 
describing the ${\cal N}=(1,1)$ reduction on $S^3$,
the potential (\ref{NVD3}) depends exclusively on the block $M^{AB}$.
In the parametrization (\ref{paramM}), the potential is thus independent of the scalars $\phi_{A\alpha}$.
Explicitly, it takes the form
\bea
   V_{(A)} &=&    2\,\alpha^2 \, e^{4\varphi} + \frac12\,e^{2\varphi}\left(
   2 \, \tilde{m}^{AB} \, \tilde{m}^{AB}  -\tilde{m}^{AA} \, \tilde{m}^{BB} \right)
  \;,
   \label{pot11}
\eea
where we have defined
\bea
m^{AB} &=& \tilde{m}^{AB}\,e^{\varphi}
\;,
\label{mm20}
\eea
with ${\rm det}\,\tilde{m}^{AB}=1$\,.
This precisely agrees with the result of \cite{Deger:2014ofa} as required by consistency
since the additional scalar fields $\phi_{A\alpha}$ do not show up in the potential.
Note that rescaling $\varphi\rightarrow\varphi-{\rm log}\,|\alpha|$ turns
the constant $\alpha$ into a global scaling factor in front of the potential,
which is thus irrelevant for the existence of stationary points.
Depending on the sign of $\alpha$ there are however two different fermionic completions
of the theory.

In contrast, for the embedding tensor (B) from (\ref{embeddingAB}), 
describing the ${\cal N}=(2,0)$ reduction on $S^3$,
the potential after
some computation takes the form
\bea
   V_{(B)}    &=& ({\rm det} \, m^{CD}) \left( 
                         2 \Big( \alpha\,  -\frac1{4} \, (\phi\phi)_{AA} \Big)^2
                         + m_{AB} \Big( m_{AB} +\frac1{2} \, (\phi\phi)_{AB} \Big)
                         -\frac1{2} \, m_{AA} \, m_{BB} \right)
                         \nonumber 
                         \\
  &=&   2\,\alpha^2 \,  e^{4\varphi}\,
	     + \frac1{2} e^{2\varphi}\,  (2\,\tilde m_{AB}\tilde m_{AB}-\tilde m_{AA}\tilde m_{BB})     
            \nonumber
	     \\
	     &&{}
  	     -\alpha\,   e^{4\varphi}\, \, (\phi\phi)_{CC}
     +  \frac1{2}\,e^{3\varphi}\,  \tilde m_{CD}\, (\phi\phi)_{CD}
           + \frac1{8} \, e^{4\varphi}\,
            (\phi\phi)_{CC}  (\phi\phi)_{DD}
            \;,
\label{pot20}
\eea
where again we used the parametrization (\ref{mm20}). We note that 
for $\phi_{A\alpha}=0$, this expression coincides with the potential (\ref{pot11})
upon flipping
\bea
\tilde{m}^{AB}&\longleftrightarrow&\tilde{m}_{AB}
\;.
\label{flipM}
\eea
This is consistent with the fact that upon setting the $\phi_{A\alpha}$ fields to zero, both the 
${\cal N}=(1,1)$ and the ${\cal N}=(2,0)$ theories reduce to the same ${\cal N}=(1,0)$ theory 
in six dimensions, which gives rise to the potential computed in \cite{Deger:2014ofa}.
Again, the constant $\alpha$ can be absorbed by shifting $\varphi$ together with a rescaling 
of $\phi_{A\alpha}$. However, the presence of a term linear in $\alpha$ implies that there are 
two inequivalent theories depending on the sign of $\alpha$ which cannot be absorbed into 
a field redefinition. In the following, we will adopt the normalization $|\alpha|=1$\,.

\subsection{Extrema of the scalar potential}

In this section, we derive the full set of extremal points of the scalar potentials (\ref{pot11}) and (\ref{pot20}).
Since (\ref{pot11}) sits within (\ref{pot20}) as a truncation $\phi_{A\alpha}=0$, it will be sufficient to analyze 
the extremal points of the latter. Below we will then uplift some of these extremal points to solutions
of the six-dimensional supergravities.

Variation of (\ref{pot20}) with respect to the scalar field $\varphi$ yields the condition
\bea
\delta_\varphi V_{(B)}  &\stackrel{!}{=}0&
\label{varphivar}
\\
\Longrightarrow
\qquad
0 &=&  8\,\alpha^2 \, 
	     +  e^{-2\varphi}\,  (2\,\tilde m_{AB}\tilde m_{AB}-\tilde m_{AA}\tilde m_{BB})     
	 %   \nonumber \\   &&{}
  	     -\alpha  \, (\phi\phi)_{CC}
           - \frac1{4} \,
            (\phi\phi)_{CC}  (\phi\phi)_{DD}
            \;.
            \nonumber
\eea
Next, let us consider the variation with respect to the $\phi_{A\alpha}$, such that  
$\delta_\Sigma (\phi\phi)_{AB} = \Sigma_{A\alpha} \phi_{B\alpha}+\Sigma_{B\alpha} \phi_{A\alpha}$.
Variation of the potential yields
\bea
\delta_\Sigma V_{(B)} &=&\Sigma_{A\alpha}\left(
  -2\,\alpha\,   e^{4\varphi}\, \phi_{A\alpha}
     +e^{3\varphi}\,  \tilde m_{AB}\,  \phi_{B\alpha}
           + \frac1{2} \, e^{4\varphi}\,
             (\phi\phi)_{DD}\, \phi_{A\alpha}
\right)
\;,
\eea
such that extremization leads to the eigenvector equation
\bea
\delta_\Sigma V_{(B)} \stackrel{!}{=}0
&\Longrightarrow&
  \tilde m_{AB}\,  \phi_{B\alpha} ~=~
 e^{\varphi} \left(2\,\alpha\,  -\frac1{2} \, 
             (\phi\phi)_{DD}\right) 
             \phi_{A\alpha}
\;.
\label{eigen}
\eea
Finally, variation with respect to the ${\rm SL}(4)$ scalars according to 
$\delta_\Lambda \tilde m_{AB}=2\,\Lambda_{(A}{}^C \tilde m_{B)C}$
with traceless $\Lambda_A{}^B$ gives rise to
\bea
\delta_\Lambda V_{(B)} \stackrel{!}{=}0
&\Longrightarrow&
0 ~=~
\left(
4\, \tilde m_{CB}\tilde m_{AB}-2\,  \tilde m_{CA}\tilde m_{BB}
   + e^{\varphi}\,  \tilde m_{CD}\, (\phi\phi)_{AD}
\right)
\Lambda_A{}^C
\;.
\label{varLV}
\eea
Upon reducing the last term by means of (\ref{eigen}), this equation can be solved for $(\phi\phi)_{AB}$
as
\bea
(\phi\phi)_{AB} &=&
e^{-2\varphi}\,\chi^{-1}
\left(
2\,  \tilde m_{AB}\tilde m_{CC}-
4\, \tilde m_{AC}\tilde m_{BC}
-\frac12\,\delta_{AB}\,  \tilde m_{CC}\tilde m_{DD}+
\delta_{AB}\, \tilde m_{CD}\tilde m_{CD}
\right)
\nonumber\\
&&{}
+\frac14\,\delta_{AB}\,(\phi\phi)_{CC}
\;,
\eea             
with $\chi=\left(2\,\alpha\,  -\frac1{2} \, (\phi\phi)_{DD}\right)$.
Plugging this expression back into the eigenvector equation (\ref{eigen})
eventually implies
\bea
 0 
 &=&
 e^{-2\varphi}\,\chi^{-1}
\left(
2\,  \tilde m_{AB}\, \tilde m_{BC}\tilde m_{DD}-
4\, \tilde m_{AB}\, \tilde m_{BD}\tilde m_{CD}
-\frac12\, \tilde m_{AC}\,\tilde m_{EE}\tilde m_{DD}+
 \tilde m_{AC}\, \tilde m_{DE}\tilde m_{DE}
\right)
\nonumber\\
&&{}
-e^{-\varphi}\,
\left(
2\,  \tilde m_{AC}\tilde m_{DD}-
4\, \tilde m_{AD}\tilde m_{CD}
-\frac12\,\delta_{AC}\,  \tilde m_{EE}\tilde m_{DD}+
\delta_{AC}\, \tilde m_{DE}\tilde m_{DE}
\right)
\nonumber\\
&&{}
+\frac14\, \tilde m_{AC}\,(\phi\phi)_{DD}
-\frac14\,e^\varphi\,\chi\,\delta_{AC}\,(\phi\phi)_{DD}
\;.
\label{finalequ}
\eea
The conditions for stationary points thus boil down to solving equations
(\ref{varphivar}) and (\ref{finalequ}). 
The value of the potential at an extremal point is computed by evaluating (\ref{pot20}) 
using (\ref{varphivar}) and (\ref{eigen})
\bea
   V_{(B),0}    &=&    -\alpha \,  e^{4\varphi}\, 
                               \left( 2 \, \alpha - \frac12 \, (\phi\phi)_{CC} \right)
\;,
\eea
which corresponds to a three-dimensional AdS length
\bea
\ell^2&=& \frac{2}{|V_0|}
\;,
\qquad
\mathring{R}_{\mu\nu} ~=~ -\frac{2}{\ell^2}\,\mathring{g}_{\mu\nu}
\;,
\eea
with $\mathring{g}_{\mu\nu}$ denoting the AdS$_3$ metric.

Let us first consider the sector $\phi_{A\alpha}=0$, which is a consistent truncation of the potential (\ref{pot20})
and contains the stationary points common to (\ref{pot11}) and (\ref{pot20}). In this case, equation (\ref{eigen}) is
trivially satisfied. Solutions of the remaining equations (\ref{varphivar}), (\ref{varLV}) are most conveniently found
in a basis in which $\tilde{M}_{AB}$ is diagonal. Inspection reveals a one-parameter family of solutions given by
\bea
m^{AB} &=&{\rm diag}\left\{e^\eta,e^\eta,e^{-\eta},e^{-\eta}\right\}\;,\quad
\phi_{A\alpha} ~=~0
\;.
\label{family}
\eea
The existence of this flat direction in the scalar potential has already been noted in~\cite{Deger:2019jtl}.
The potential for these families remains fixed at $V_{(B),0}=-2$, and the
scalar spectrum is given by
\bea
 m^2\ell^2&:& 
0\;[5]\;,\quad
8\;[1]\;,\quad
4\,e^{2\eta}-4\;[2]\;,\quad
4\,e^{-2\eta}-4\;[2]\;,
\eea
completed by
\bea
\left\{
\begin{array}{rl}
0\;\;[16] &:\; \mbox{potential (\ref{pot11})} \\
e^{2\eta}-2\,e^{\eta}\,({\rm sgn}\,\alpha)\;[8]\;,\quad
e^{-2\eta}-2\,e^{-\eta}\,({\rm sgn}\,\alpha)\;[8] &:\; \mbox{potential (\ref{pot20})}
\end{array}
\right\}
\;,
\eea
for the different potentials. These spectra are stable 
(in the Breitenlohner-Freedman sense $m^2\ell^2\ge-1$ \cite{Breitenlohner:1982jf}) for 
\bea
\frac12\,{\sqrt{3}} ~\le~e^\eta~\le~ \frac23\,{\sqrt{3}}
\;.
\eea
The vector spectrum is given by
\bea
m\ell&:& 
\pm2\;[1+1]\;,\quad
1\pm\sqrt{2\,{\rm cosh}(2\eta) - 1}\;[2+2]\;,\quad
-1\pm\sqrt{2\,{\rm cosh}(2\eta) - 1}\;[2+2]\;,\quad
\eea
reflecting the unbroken ${\rm SO}(2)\times {\rm SO}(2)\subset {\rm SO}(4)$\,.
Finally, the gravitino spectrum is given by
\bea
m\ell&:& 
\pm\frac12\left(2\,{\rm cosh}\,\eta +({\rm sgn}\,\alpha)\right)\;[4+4]\;,
\eea
showing that only for $\alpha=-1$,
the vacuum at $\eta=0$ is supersymmetric, preserving ${\cal N}=(4,4)$ supersymmetry.
This corresponds to the six-dimensional supersymmetric background AdS$_3\times S^3$\,.
The $\alpha=+1$ solution is not supersymmetric, but may correspond to a supersymmetric
solution in an ${\cal N}=(1,0)$ theory coupled to tensor multiplets.

The potential (\ref{pot20}) allows for additional stationary points with $\phi_{A\alpha}\not=0$\,. 
In this case, the remaining equations (\ref{varphivar}), (\ref{finalequ}) again are most conveniently solved 
in a basis in which $\tilde{m}_{AB}$ is diagonal, where we find 4 discrete solutions. 
They all necessitate positive $\alpha=+1$ with the potential taking the values
\bea
V_{(B),0} &=& 
\left\{
\begin{array}{rl}
-\frac{27}{8}& (i)\\
-\frac{8788}{3125}&(ii)\\
-4&(iii)\\
-\frac{25}{8}&(iv)
\end{array}
\right.
\;.
\label{stationary}
\eea
All these stationary points fully break supersymmetry and SO(4) gauge symmetry, and
they all contain unstable scalars with masses below the
Breitenlohner-Freedman bound $m^2\ell^2=-1$. 
For later checks, let us only note the location of solution $(i)$
\bea
m^{AB} &=& \mathfrak{m}\,\delta^{AB} ~=~ \frac3{2}\,\delta^{AB}\;,\quad
\phi_{A\alpha} ~=~
\frac{\sqrt{2}}{\sqrt{3}}\,\delta_{A\alpha}
\;,
\label{soli}
\eea
with the scalar mass spectrum given by
\bea
 m^2\ell^2& :&\quad
\frac23\pm 2\;[9+9]\;,\quad
6\pm2\;[1+1]\;\;,\quad
0\;[6]\;,
\eea
in units of the AdS length $\ell=\frac{4}{3\sqrt{3}}$\;.

%%%%%%%%%%%%%%%%%%%%%%%%%%%%%%%

\section{${\rm SO}(8,4)$ exceptional field theory}
\label{sec:ExFT}

%%%%%%%%%%%%%%%%%%%%%%%%%%%%%%%

In this section, we review the structure of ${\rm SO}(8,4)$ ExFT, 
constructed in \cite{Hohm:2017wtr}, to which we refer for details. 
This theory provides the manifestly duality covariant formulation
of the 6D supergravity theories relevant for our consistent truncations.
We discuss the inequivalent solution to its section constraints and establish the dictionary 
of the ExFT fields to the 6D fields of ${\cal N}=(1,1)$ and ${\cal N}=(2,0)$ supergravity 
theories, respectively.

\subsection{Lagrangian}

Similar to the three-dimensional supergravities reviewed in section~\ref{sec:3D}, ${\rm SO}(8,4)$ ExFT is based
on the coset space (\ref{coset}) which we parametrize by a symmetric positive definite matrix
${\cal M}_{MN}$\,. In contrast to the matrix of (\ref{paramM}), this matrix depends not
only on three external coordinates $x^\mu$, but in addition on (${\dim}\,{\rm SO}(8,4)$) coordinates
$Y^{MN}$ with the latter dependence strongly constrained by the section conditions
\bea
\partial_{[MN}\otimes \partial_{KL]} \ = \  0 ~=~ \eta^{NK}\,\partial_{MN}\otimes \partial_{KL}
\;, 
\label{section}
\eea
which restrict the fields to live on sections of dimension three (at most). Depending on the 
choice of these sections, the theory describes the 6D ${\cal N}=(1,1)$ or ${\cal N}=(2,0)$ supergravity, respectively. 
The theory is invariant under generalized internal diffeomorphisms, acting as
\bea
{\cal L}_{\Lambda, \Sigma} {\cal M}^{MN} &=& 
\Lambda^{KL}\,\partial_{KL} {\cal M}^{MN}
+ 
4\left(
\partial^{K(M} \Lambda_{KL} 
-\partial_{KL} \Lambda^{K(M} 
+ 2\,\Sigma^{(M}{}_{L}
\right)
{\cal M}^{N)L}
\;,
\label{gen_diff}
\eea
on the scalar matrix. Here, the gauge parameters $\Sigma_{MN}$ are subject to algebraic constraints
analogous to (\ref{section}), i.e.
\bea
\Sigma_{[MN}\otimes \Sigma_{KL]} \ = \  0 ~=~ \eta^{NK}\,\Sigma_{MN}\otimes \Sigma_{KL}
\;,
\label{sectionS}
\eea
as well as compatibility with the partial derivatives as
\bea
\Sigma_{[MN}\otimes \partial_{KL]} \ = \  0 ~=~ \eta^{NK}\,\Sigma_{MN}\otimes \partial_{KL}
\;.
\label{sectionS0}
\eea
Invariance under local internal diffeomorphisms (\ref{gen_diff}) is ensured by minimal couplings to
gauge fields $({\cal A}_\mu{}^{MN}, {\cal B}_{\mu\,MN})$ via covariant external derivatives
\bea
D_{\mu} &=& \partial_\mu - {\cal L}_{{\cal A}_\mu,{\cal B}_\mu}
\;.
\label{covd}
\eea
The full Lagrangian is  given by
\bea
{\cal L} &=& 
{\cal L}_{\rm EH} +  {\cal L}_{\rm kin} + {\cal L}_{\rm CS} -\sqrt{-g}\,V_{\rm ExFT}
\;,
\label{action0}
\eea
each term being separately invariant under generalized internal diffeomorphisms (\ref{gen_diff}). 
The modified Einstein-Hilbert term and the scalar kinetic term have the form
\bea
{\cal L}_{\rm EH}&=&  \sqrt{-g}\,e_a{}^\mu e_b{}^\nu \left(
{R}_{\mu\nu}{}^{ab}+ { F}_{\mu\nu}{}^{M N}e^{a\,\rho}\partial_{MN}e_{\rho}{}^{b}
\right)
~\equiv~  \sqrt{-g}\,\hat{R}\nonumber\;,\\
{\cal L}_{\rm kin}&=&\frac{1}{8}\,   \sqrt{-g}\,\,g^{\mu\nu}\,D_{\mu}{\cal M}^{MN} D_\nu {\cal M}_{MN},
\label{LEH}
\eea
with the covariant derivatives (\ref{covd}), Yang-Mills field strength $F_{\mu\nu}{}^{MN}$, and the Riemann
tensor ${R}_{\mu\nu}{}^{ab}$ computed from the external vielbein $e_\mu{}^a$
with derivatives covariantized under internal diffeomorphisms under which
$e_\mu{}^a$ transforms as a scalar density (of weight $\lambda=1$).
The gauge fields couple with a Chern-Simons term that takes the explicit form
\bea
{\cal L}_{\rm CS} &=&
\sqrt{2}\,\varepsilon^{\mu\nu\rho}\,\Big( 
{F}_{\mu\nu}{}^{MN} {\cal B}_{\rho\,MN}
+ \partial_\mu {\cal A}_{\nu\,N}{}^K \partial_{K M} {\cal A}_\rho{}^{MN}
-\frac23\,
\partial_{MN}\partial_{KL} {\cal A}_{\mu}{}^{KP} {\cal A}_\nu{}^{MN} {\cal A}_{\rho\,P}{}^{L}
\nonumber\\
&&{}
\qquad\qquad
+\frac23\, {\cal A}_{\mu}{}^{LN} \partial_{MN} {\cal A}_\nu{}^{M}{}_{P} \partial_{KL} {\cal A}_\rho{}^{PK}
-\frac43\,{\cal A}_{\mu}{}^{LN} \partial_{MP} {\cal A}_\nu{}^{M}{}_{N} \partial_{KL} {\cal A}_\rho{}^{PK}\Big) 
\;.\quad
\label{CS}
\eea
Finally, the last term in (\ref{action0}) carries only internal derivatives $\partial_{MN}$ and is given by
\bea
V_{\rm ExFT} &\equiv&
-\frac18\,
{\cal M}^{KP}{\cal M}^{LQ}\,
\partial_{KL} {\cal M}_{MN}\,
\partial_{PQ} {\cal M}^{MN}
-\frac12\,\partial_{MK} {\cal M}^{NP}
\partial_{NL} {\cal M}^{MQ}\,{\cal M}^{KL}{\cal M}_{PQ} 
 \nonumber\\
 &&{}
-\frac14\,
\partial_{MN} {\cal M}^{PK}\,
\partial_{KL} {\cal M}^{QM}\,{\cal M}_{P}{}^L{\cal M}_{Q}{}^N
\nonumber
+2\,\partial_{MK} {\cal M}^{NK}\,\partial_{NL} {\cal M}^{ML}
 \nonumber\\
 &&{}
-g^{-1}\partial_{MN} g\,\partial_{KL}{\cal M}^{MK}{\cal M}^{NL}
 -\frac{1}{4}  \,{\cal M}^{MK}{\cal M}^{NL}\,g^{-2}\partial_{MN}g\,\partial_{KL}g
\nonumber\\
 &&{}
  -\frac{1}{4}\,{\cal M}^{MK}{\cal M}^{NL}\,\partial_{MN}g^{\mu\nu}\partial_{KL} g_{\mu\nu}
  \;.
  \label{full_potential}
\eea
Depending on the solution of the section constraints (\ref{section}), the action
(\ref{action0}) describes 6D ${\cal N}=(1,1)$ or ${\cal N}=(2,0)$ supergravity. 
In the next two subsections, we review the two inequivalent solutions to the
section constraints and the associated dictionaries of the ExFT fields into
the 6D supergravity fields.

%%%%%%%%%%%%%%%%%%%%%%%%%%%%%%%

\subsection{${\cal N}=(1,1)$ solution of section constraint}
\label{subsec:N11ssc}

%%%%%%%%%%%%%%%%%%%%%%%%%%%%%%%

Consider the decomposition of ${\rm SO}(8,4)$ under its subgroup
\bea
{\rm GL}(3)\times {\rm SO}(1,1)\times {\rm SO}(4)\subset  {\rm SO}(4,4)\times {\rm SO}(4)\subset {\rm SO}(8,4)\;,
\eea 
such that the fundamental vector of ${\rm SO}(8,4)$ decomposes as
\bea
\left\{ V^M \right\}&\longrightarrow&
\left\{ (V^i)_{(-1)}, (V_{i})_{(+1)}, (V^0)_{(-3)}, (V_0)_{(+3)} , (V^\alpha)_{(0)} \right\}
\;,
\label{basis11}
\eea
where subscripts refer to the sum of the ${\rm GL}(1)\subset {\rm GL}(3)$ charge and the ${\rm SO}(1,1)$ charge,
defining the grading associated to the higher-dimensional origin of these fields. 
Here $i=1, 2, 3$ and $\alpha=1, \dots, 4$.
The invariant tensor $\eta_{MN}$ decomposes accordingly
\bea
\eta_{MN} &=& 
\begin{pmatrix}
0 & \delta_i{}^j &0 & 0 &0\\
 \delta_j{}^i &0 & 0 & 0& 0\\
0 &0 & 0& 1 & 0 \\
0 & 0&1& 0&0\\
0&0&0&0& -\delta_{\alpha\beta}
\end{pmatrix}
\;.
\eea
The ${\cal N}=(1,1)$ solution to the section constraints (\ref{section}) is given by 
decomposing coordinates $Y^{MN}$ according to (\ref{basis11}) and restricting the 
internal coordinate dependence of all fields to the coordinates $Y^{0i}$, such that
the only non-vanishing internal derivatives are
\bea
\partial_i \equiv \frac1{\sqrt{2}}\,\partial_{0i} \;,
\label{der11}
\eea
providing a solution to (\ref{section}).
Breaking the ExFT fields according to (\ref{basis11}), then 
matches the field content of the 6D ${\cal N}=(1,1)$
supergravity which, in addition to the metric and the dilaton,
contains 4 vector fields and a (non-chiral) 2-form gauge field.

Specifically, the ExFT vector fields transform in the adjoint representation of ${\rm SO}(8,4)$. 
Under (\ref{basis11}), they decompose into
\bea
{\cal A}_\mu{}^{MN} &\longrightarrow& 
\left\{
\begin{array}{rclcl}
-4&:&{\cal A}_{\mu}{}^{0i}&\subset& \mbox{6d metric} \\
-3&:&{\cal A}_{\mu}{}^{0\alpha}&\subset& \mbox{6d vectors} \\
-2&:& {\cal A}_{\mu\,i}{}^0, {\cal A}_{\mu}{}^{ij}  &\subset& \mbox{6d 2-form and its dual} \\
-1&:&{\cal A}_{\mu}{}^{i\alpha}&\subset& \mbox{6d dual 3-form} \\
0 &:& {\cal A}_{\mu\,i}{}^j, {\cal A}_{\mu\,0}{}^{0}, {\cal A}_\mu{}^{\alpha\beta}  &\subset& \mbox{6d dual graviton, etc.} \\
+1 &:& \dots
\end{array}
\right.
\;,
\label{vectorsN11}
\eea
allowing to identify the higher-dimensional origin of the various components.
The fields of positive grading do not enter the action (\ref{action0}).
Similarly, one decomposes the scalar fields, parametrizing the coset ${\rm SO}(8,4)\big/\left(
{\rm SO}(8)\times {\rm SO}(4)
\right)$ into
\bea
&&
\left\{
\begin{array}{rclcl}
+4&:&\phi_{0i}&\subset& \mbox{6d dual graviton} \\
+3&:&\phi_{\alpha 0}&\subset& \mbox{6d dual 3-form} \\
+2&:& \phi^{i}{}_0, \phi_{ij} &\subset& \mbox{6d 2-form and its dual} \\
+1&:& \phi_{i\alpha} &\subset& \mbox{6d vectors} \\
0 &:& g_{ij}, \varphi   &\subset& \mbox{6d metric and dilaton} \\
\end{array}
\right.
\;.
\eea
In order to identify their location within the scalar matrix ${\cal M}_{MN}$, 
it is useful to determine the action of a generalized diffeomorphism (\ref{gen_diff})
on the various components of the matrix ${\cal M}^{MN}$.  In the
decomposition (\ref{basis11}) this gives particularly neat expressions when acting on some
specific combinations
\bea
{\cal L}_{\Lambda, \Sigma} {\cal M}^{00} &=&
L_\lambda {\cal M}^{00} - 2\,(\partial_k \lambda^k) {\cal M}^{00}
\;,\nonumber\\[2ex]
{\cal L}_{\Lambda, \Sigma} {\cal M}^{0i} &=&
L_\lambda {\cal M}^{0i}  - (\partial_k \lambda^k) {\cal M}^{0i}
- \varepsilon^{ijk}\,\partial_{j}\xi_k  {\cal M}^{00}
\;,\nonumber\\[2ex]
{\cal L}_{\Lambda, \Sigma}
\left( {\cal M}^{00} {\cal M}^{ij}-{\cal M}^{0i}{\cal M}^{0j}\right)
&=&
L_\lambda \left( {\cal M}^{00} {\cal M}^{ij}-{\cal M}^{0i}{\cal M}^{0j}\right)
\nonumber\\
&&{}
- 2\,(\partial_m \lambda^m)   \left( {\cal M}^{00} {\cal M}^{ij}-{\cal M}^{0i}{\cal M}^{0j}\right)
\;,\nonumber\\[2ex]
{\cal L}_{\Lambda, \Sigma}
\left({\cal M}^{00}{\cal M}^{\alpha i}- {\cal M}^{0\alpha} {\cal M}^{0i}\right)
&=&
L_\lambda \left({\cal M}^{00}{\cal M}^{\alpha i}- {\cal M}^{0\alpha} {\cal M}^{0i}\right)
+
\partial_{j} \Lambda^{\alpha} \left(
{\cal M}^{00} {\cal M}^{ij} 
-   {\cal M}^{0i} {\cal M}^{0j} \right)
\nonumber\\
&&{}
- 2\,(\partial_m \lambda^m)  \left({\cal M}^{00}{\cal M}^{\alpha i}- {\cal M}^{0\alpha} {\cal M}^{0i}\right)
\;,\nonumber\\[2ex]
{\cal L}_{\Lambda, \Sigma}
\left({\cal M}^{00}{\cal M}^{i}{}_j- {\cal M}^{0i} {\cal M}^{0}{}_{j}\right)
&=&
L_\lambda \left({\cal M}^{00}{\cal M}^{i}{}_j- {\cal M}^{0i} {\cal M}^{0}{}_{j}\right)
+\partial_{j} \Lambda^{\alpha} \left(  {\cal M}^{00} {\cal M}^{\alpha i} - {\cal M}^{0\alpha} {\cal M}^{0i} \right)
\nonumber\\
&&{}
+\left(
\partial_{k} \tilde\xi_{j}
-\partial_{j}  \tilde\xi_{k}  \right)
\left(  {\cal M}^{00}  {\cal M}^{ik} -  {\cal M}^{0i} {\cal M}^{0k}\right)
\nonumber\\
&&{}
- 2\,(\partial_m \lambda^m)   \left({\cal M}^{00}{\cal M}^{i}{}_j- {\cal M}^{0i} {\cal M}^{0}{}_{j}\right)
\;,
\eea
where we have redefined the gauge parameters as
\bea
&&
\Lambda^{0i}= \frac1{\sqrt{2}}\,\lambda^i
\;,\quad
 \Lambda^{ij}=
 \frac1{\sqrt{2}}\,\varepsilon^{ijk}\,\xi_k
\;,\quad
 \Lambda_i{}^0=
 \frac1{\sqrt{2}}\,\tilde\xi_i
\;,\nonumber\\
&& \Lambda^{\alpha 0} =
\frac1{\sqrt{2}}\,\Lambda^\alpha
\;,\quad
\Lambda^{\alpha i} =
\frac1{2\sqrt{2}}\,\varepsilon^{ijk}\,\Lambda_{jk}{}^\alpha
\;,
\label{gauge_param_11}
\eea
with the totally antisymmetric $\varepsilon^{ijk}$.
Here, $L_\lambda$ denotes the standard Lie derivative along the vector field $\lambda^k$.
Identifying the higher-dimensional origin of the gauge parameters among internal 6D
diffeomorphisms and gauge transformations according to the identification of the vector fields (\ref{vectorsN11})
then allows to read off the dictionary between 
the components of ${\cal M}^{MN}$ and the internal components of the 6D fields
\bea
{\cal M}^{00} &=& g^{-1}\,e^\phi
\;,\nonumber\\
{\cal M}^{0i} &=& -\frac12\,{\cal M}^{00}\,\varepsilon^{ijk}\,B_{jk}
\;,
\nonumber\\
{\cal M}^{00} {\cal M}^{ij}-{\cal M}^{0i}{\cal M}^{0j} &=& g^{-1}\,g^{ij}
\;,\nonumber\\
{\cal M}^{00}{\cal M}^{\alpha i}- {\cal M}^{0\alpha} {\cal M}^{0i} &=& g^{-1}\,g^{ij}\,A_j{}^\alpha
\;,\nonumber\\
{\cal M}^{00}{\cal M}^{i}{}_j- {\cal M}^{0i} {\cal M}^{0}{}_{j} &=&
g^{-1}\,g^{ik} \,\tilde B_{kj} 
+ 
\frac12\,g^{-1}g^{ik}\,A_k{}^\alpha A_j{}^\alpha
\;.
\label{compM11}
\eea
The dictionary is such that the generalized diffeomorphisms (\ref{gen_diff})
reproduce the gauge transformations
\bea
\delta A_i{}^\alpha &=& \partial_i \Lambda^\alpha\;,\nonumber\\
\delta B_{ij} &=&
2\,\partial_{[i}  \xi_{j]} \;,\nonumber\\
\delta\tilde B_{ij} &=&
2\,\partial_{[i} \tilde \xi_{j]} 
+ \partial_{[i} \Lambda{}^\alpha A_{j]}{}^\alpha
~=~
2\,\partial_{[i} \left(\tilde \xi_{j]}+\frac12\, \Lambda{}^\alpha A_{j]}{}^\alpha \right)
- \frac12\,  \Lambda{}^\alpha F_{ij}{}^\alpha
\;,
\eea
of the 6D vector fields $A_i{}^\alpha$, 2-form $B_{ij}$ and its dual $\tilde{B}_{ij}$\,.
Finally, using the dictionary (\ref{compM11}), we may also consider
\bea
{\cal L}_{\Lambda, \Sigma} \left(  ({\cal M}^{00})^{-1} {\cal M}^{0\alpha} \right) 
&=&
L_\lambda  \left(  ({\cal M}^{00})^{-1} {\cal M}^{0\alpha} \right)   
+ (\partial_k \lambda^k)  \left(  ({\cal M}^{00})^{-1} {\cal M}^{0\alpha} \right) 
\nonumber\\
&&{}
- \frac12\,\varepsilon^{ijk} \partial_{i} \Lambda_{jk}{}^{\alpha} 
-\frac12\,\partial_{i} \Lambda^{\alpha} \, \varepsilon^{ijk}\,B_{jk}
\;,
\eea
from which we infer the identification
\bea
({\cal M}^{00})^{-1} {\cal M}^{0\alpha} &=&
- \frac16\,\varepsilon^{ijk}\,a_{ijk}{}^\alpha
\;,
\eea
with the scalars $a_{ijk}{}^\alpha$ from the dual 3-forms in six dimensions.
These transform as
\bea
\delta a_{ijk}{}^\alpha &=&
3\,\partial_{[i} \Lambda_{jk]}{}^{\alpha} 
+3\,B_{[ij}\,\partial_{k]} \Lambda^{\alpha} 
\eea
under 6D gauge transformations.

%%%%%%%%%%%%%%%%%%%%%%%%%%%%%%%

\subsection{${\cal N}=(2,0)$ solution of section constraint}
\label{subsec:N20ssc}

%%%%%%%%%%%%%%%%%%%%%%%%%%%%%%%

Consider the decomposition of ${\rm SO}(8,4)$ under its subgroup
${\rm GL}(3)\times {\rm SO}(1,5)$ such that the fundamental ${\rm SO}(8,4)$ vector decomposes as
\bea
\left\{ V^M \right\}&\longrightarrow&
\left\{ V_{i\,(-2)}, V^i{}_{(+2)}, V^0_{(0)} , V^a_{(0)} \right\}
\;,
\label{basis1}
\eea
with subscripts referring to ${\rm GL}(1)$ charges. Here $i=1, 2, 3$ and $a=1, \dots, 5$.
The invariant metric $\eta_{MN}$ decomposes as
\bea
\eta_{MN} &=& 
\begin{pmatrix}
0 & \delta^i{}_j &0 & 0\\
 \delta^j{}_i &0 & 0 & 0\\
0 &0 & 1 & 0 \\
0 & 0&0& -\delta_{ab}
\end{pmatrix}
\;.
\eea
Later on, we will also decompose $a\longrightarrow\{\bar0,\alpha\}$\,.
The ${\cal N}=(2,0)$ solution to the section constraints (\ref{section}) is given by 
decomposing coordinates $Y^{MN}$ according to (\ref{basis1}) and restricting the 
internal coordinate dependence of all fields to the coordinates $Y_{ij}$, such that
the only non-vanishing internal derivatives are
\bea
\partial_i \equiv \frac12\,\varepsilon_{ijk}\,\partial^{jk} \;,
\label{der20}
\eea
providing a solution to (\ref{section}).
Breaking the ExFT fields according to (\ref{basis1}) then matches the field content of the 6D ${\cal N}=(2,0)$
supergravity coupled to a tensor multiplet, which contains 5 selfdual and 1 anti-selfdual 
2-form gauge fields, together with 5 scalar fields parametrizing the coset space 
${\rm SO}(1,5)/{\rm SO}(5)$\,.

Specifically, the ExFT vector fields transform in the adjoint representation of ${\rm SO}(8,4)$. 
Under (\ref{basis1}), they decompose into
\bea
{\cal A}_\mu{}^{MN} &\longrightarrow& 
\left\{
\begin{array}{rclcl}
-4&:&{\cal A}_{\mu\,ij}&\subset& \mbox{6d metric} \\
-2&:& {\cal A}_{\mu\,i}{}^{a}, {\cal A}_{\mu\,i}{}^{0} &\subset& \mbox{6d 2-forms} \\
0 &:& {\cal A}_{\mu\,i}{}^j, {\cal A}_\mu{}^{ab}, {\cal A}_\mu{}^{a0}  &\subset& \mbox{6d dual graviton, etc.} \\
2 &:& \dots
\end{array}
\right.
\label{vectorsN20}
\eea
allowing to identify the higher-dimensional origin of the various components.
The fields of positive grading do not enter the action (\ref{action0}).
Similarly, one decomposes the scalar fields, parametrizing the coset ${\rm SO}(8,4)\big/\left(
{\rm SO}(8)\times {\rm SO}(4)
\right)$ into
\bea
&&
\left\{
\begin{array}{rclcl}
+4&:&\phi^{ij}&\subset& \mbox{6d dual graviton} \\
+2&:& \phi_{a}{}^{i}, \phi_{0}{}^{i} &\subset& \mbox{6d 2-forms} \\
0 &:& m_{ij}, \phi, m_{\bar a\bar b}  &\subset& \mbox{6d metric and scalars} \\
\end{array}
\right.
\;,
\eea
where $\bar a=0, .., 5$ and $m_{ij}$ and $m_{\bar a\bar b}$ parametrize the 
coset spaces ${\rm SL}(3)/{\rm SO}(3)$ and ${\rm SO}(1,5)/{\rm SO}(5)$, respectively.
For the latter, we have the invariant tensor
\bea
\eta_{\bar a \bar b} &=& 
\begin{pmatrix}
1 & 0\\
0 & -\delta_{ab}
\end{pmatrix}
\;. \label{mink15}
\eea
In order to identify the precise location of scalar fields within the scalar matrix ${\cal M}_{MN}$, 
it is useful to determine the action of a generalized diffeomorphism (\ref{gen_diff})
on the various components of the matrix ${\cal M}^{MN}$.  In the
decomposition (\ref{basis1}) this  takes the form
\bea
{\cal L}_{\Lambda, \Sigma} {\cal M}_{ij} &=&
L_\lambda {\cal M}_{ij} - 2\,(\partial_k \lambda^k) {\cal M}_{ij}
\;,\nonumber\\
{\cal L}_{\Lambda, \Sigma} {\cal M}_i{}^{\bar a} &=&
L_\lambda {\cal M}^{i\bar a}  - (\partial_k \lambda^k) {\cal M}_i{}^{\bar a} 
+2\,\varepsilon^{kmn}\,{\cal M}_{ik}\,\partial_m \Lambda^{\bar{a}}{}_{n}
\;,\nonumber\\
{\cal L}_{\Lambda, \Sigma} {\cal M}^{\bar a\bar b} &=&
L_\lambda\,{\cal M}^{\bar a\bar b} +4\,
\varepsilon^{ijk}\,\,{\cal M}_k{}^{(\bar a} \partial_i \Lambda^{\bar{b})}{}_{j}
\;,
\eea
with the gauge parameter relabelled as
\bea
\Lambda_{ij} &\equiv&\frac12\,\varepsilon_{ijk}\,\lambda^k
\;.
\eea
These let us infer the dictionary
\bea
{\cal M}_{ij} &=& g^{-1}\,g_{ij}
\;,\nonumber\\
{\cal M}_i{}^{\bar a} &=&  g^{-1}\,g_{ij}\,\varepsilon^{jkl}\,B_{kl}{}^{\bar a}
\;,\nonumber\\
{\cal M}^{\bar a\bar b} &=&
{\ss M}^{\bar a \bar b} + 2\, B_{ij}{}^{\bar a}\,B_{kl}{}^{\bar b}\,g^{ik}g^{jl}
~=~{\ss M}^{\bar a \bar b} + g\,g^{ij}\,  {\cal M}_i{}^{\bar a} {\cal M}_j{}^{\bar b}
\;,
\label{compM}
\eea
with $g\equiv{\rm det}\,g_{ij}$\,, and the components $B_{ij}{}^{\bar a}$ transform under
tensor gauge transformations as $\delta B_{ij}{}^{\bar a}= 2\,\partial_{[i}\,\Lambda^{\bar{a}}{}_{j]}$\,.

\subsection{Generalized Scherk-Schwarz reduction}

Consistent truncations of ${\rm SO}(8,4)$ ExFT can be defined by a 
generalized Scherk-Schwarz compactification ansatz~\cite{Lee:2014mla,Hohm:2014qga}, 
in which the dependence of the ExFT fields on the internal coordinates is carried by
an ${\rm SO}(8,4)$ twist matrix ${\cal U}_M{}^{\bar{N}}$ and a scalar factor $\rho$.
Specifically, the ExFT fields take the factorized form~\cite{Hohm:2017wtr}
\bea
g_{\mu\nu}(x,Y) &=& \rho(Y)^{-2}\,g_{\mu\nu}(x)\;,\nonumber\\
 {\cal M}_{MN}(x,Y) &=& 
 {\cal U}_{M}{}^{\bar{M}}(Y)\,M_{\bar M\bar N}(x)\,{\cal U}_{N}{}^{\bar{N}}(Y)
 \;,
 \nonumber\\
 {\cal A}_\mu{}^{MN}(x,Y)&=&\rho(Y)^{-1} {\cal U}^M{}_{\bar M}(Y) {\cal U}^N{}_{\bar N}(Y)A_\mu{}^{\bar M\bar N}(x)\;,\nonumber\\
 {\cal B}_{\mu\,KL}(x,Y)&=&-\frac14\,\rho(Y)^{-1}\, {\cal U}_{M\bar N}(Y)\,\partial_{KL} {\cal U}^M{}_{\bar M}(Y)\,A_\mu{}^{\bar M\bar N}(x)\;,
 \label{Scherk-Schwarz}
\eea
in terms of the $x$-dependent fields of 3D gauged supergravity reviewed in section~\ref{sec:3D} above.
The embedding tensor (\ref{embedding0}) of the 3D theory is given in terms of the twist matrix as
\bea
\theta_{\bar K\bar L\bar P\bar Q}&=&6\,\rho^{-1}\,\partial_{LP}{\cal U}_{N\,[\bar K}{\cal U}^N{}_{\bar L}
{\cal U}^L{}_{\bar P}{\cal U}^P{}_{\bar Q]}\,
\;,\nonumber\\
\theta_{\bar P\bar Q}&=&4\,\rho^{-1}\,{\cal U}^K{}_{\bar P}\,\partial_{KL}{\cal U}^L{}_{\bar Q}-\frac{\rho^{-1}}{3}\,\eta_{\bar P\bar Q}{\cal U}^{K\bar L}\partial_{KL}{\cal U}^L{}_{\bar L}
-4\,\rho^{-2}\,\partial_{\bar P\bar Q}\rho\;,\nonumber\\
\theta&=&\frac{\rho^{-1}}{3}\,{\cal U}^{K\bar L}\partial_{KL}{\cal U}^L{}_{\bar L}\;,
\label{theta_comp}
\eea
and the truncation is consistent if all three objects in (\ref{theta_comp}) are actually $Y$-independent.
Using the twist matrices from~\cite{Hohm:2017wtr} we will, in the following, use the generalized
Scherk-Schwarz ansatz in order to derive the explicit reduction formulas for the 6D consistent truncations.

%%%%%%%%%%%%%%%%%%%%%%%%%%%%%%%

\section{${\cal N}=(1,1)$ uplift formulas}
\label{sec:N11uplift}

%%%%%%%%%%%%%%%%%%%%%%%%%%%%%%%

In this and the following section, we will review from~\cite{Hohm:2017wtr}
the twist matrices inducing the embedding tensors (\ref{embeddingAB}).
Combining them with the ansatz (\ref{Scherk-Schwarz}) and the supergravity
dictionaries worked out in sections~\ref{subsec:N11ssc}, \ref{subsec:N20ssc} above,
we deduce the six-dimensional ${\cal N}=(1,1)$ and ${\cal N}=(2,0)$ reduction formulas.

\subsection{Twist matrix}
\label{subsec:N11twist}

The ${\rm SO}(8,4)$ twist matrix ${\cal U}_M{}^{\bar M}$ describing the consistent $S^3$ truncation 
of 6D ${\cal N}=(1,1)$ supergravity has been constructed in \cite{Hohm:2017wtr}.
Let us recall that the coordinates $y^i$ relevant for 6D ${\cal N}=(1,1)$ supergravity have been
identified among the $Y^{MN}$ via (\ref{der11}). The associated twist matrix is
given in terms of the elementary $S^3$ sphere harmonics ${\cal Y}^A$ 
(satisfying ${\cal Y}^A{\cal Y}^A=1$), the round $S^3$ metric 
$\mathring{g}_{ij}=\partial_i{\cal Y}^A\partial_j{\cal Y}^A$ (with determinant $\mathring{g}$),
and the vector field $\mathring\zeta^i$ defined by $\mathring{\nabla}_i \mathring\zeta^i = 1$.
By $\mathring{\omega}_{ijk}\equiv \mathring{g}^{1/2}\,\varepsilon_{ijk}$, we denote the associated volume form.
We refer to appendix~\ref{sec:appS3} for further identities among these objects.
After some rewriting, the twist matrix of \cite{Hohm:2017wtr} takes the explicit form
\bea
{\cal U}_M{}^{\bar M} &=&
\left(
\begin{array}{c:c:c}
{\cal U}_0{}^A & {\cal U}_0{}_A &0 \\ \hdashline 
{\cal U}_i{}^A & {\cal U}_i{}_A  &0 \\ \hdashline
{\cal U}^0{}^A & {\cal U}^0{}_A &0  \\ \hdashline
{\cal U}^i{}^A & {\cal U}^i{}_A & 0 \\ \hdashline
0&0& \delta_\alpha{}^\beta
\end{array}\right)
\nonumber\\
&=&
\left(
\begin{array}{c:c:c}
 \mathring{g}^{1/2}\left(
{\cal Y}^{ A} -2\,\mathring{\zeta}^i \partial_i {\cal Y}^{ A} \right)& 0& 0  \\ \hdashline 
\partial_i {\cal Y}^{ A} & 
  2\,\alpha\, \mathring{g}^{jl} \partial_l {\cal Y}^A  \,\mathring{\omega}_{ijk} \mathring{\zeta}^k
 & 0 \\ \hdashline
0 & \mathring{g}^{-1/2} {\cal Y}^A  & 0\\ \hdashline
0 & 2\, \mathring{\zeta}^i{\cal Y}^A +  \mathring{g}^{ij} \partial_j {\cal Y}^A& 0 \\ \hdashline 
0&0& \delta_\alpha{}^\beta
\end{array}\right)
\;,
\eea
in a basis where the `curved index' $M$ is decomposed according to (\ref{basis11}), 
and the `flat index' $\bar{M}$ is decomposed in the basis (\ref{GL4SO4}), suitable for
the fields of 3D supergravity.
The free parameter $\alpha$ can (up to sign) be absorbed into a shift of the 6D dilaton.

\subsection{Uplift formulas}

According to the dictionary (\ref{compM11}), the 6D dilaton is identified within the component ${\cal M}^{00}$
of the ExFT scalar matrix, such that its reduction formula is obtained via (\ref{Scherk-Schwarz}) as
\bea
 e^\phi &=& g\,{\cal M}^{00} ~=~
 g\,{\cal U}^{0}{}_A  {\cal U}^{0}{}_B m^{AB}
  ~=~ \Delta^2\,{\cal Y}^A {\cal Y}^B  m^{AB}
\;,
\label{dila0}
\eea
where we have defined the warp factor 
\bea
\Delta &\equiv& \frac{g^{1/2}}{\mathring{g}^{1/2}}
\;.
\label{warp0}
\eea
Similarly, we identify the internal components of the 6D 2-form as
\bea
-\frac12 \,\varepsilon^{ijk}\,B_{jk}&=&
({\cal M}^{00})^{-1}{\cal M}^{0i} ~=~
({\cal M}^{00})^{-1}\,\mathring{g}^{-1/2} 
\left(
2\, \mathring{\zeta}^i{\cal Y}^B +  \mathring{g}^{ij} \partial_j {\cal Y}^B \right)
{\cal Y}^A\,m^{AB}
\;,
\eea
giving rise to
\bea
B_{ij} &=&
- \mathring{\omega}_{ijk}
\left(
2\,\mathring{\zeta}^k + \frac12\, \mathring{g}^{kl} \partial_l \,
{\rm log}\left({\cal Y}^A {\cal Y}^B\,m^{AB} \right) \right)
\;.
\eea
Further computation yields the 6D internal (inverse) metric
\bea
g^{ij} &=&
g\left({\cal M}^{00} {\cal M}^{ij}-{\cal M}^{0i}{\cal M}^{0j} \right)
\nonumber\\
&=&
\Delta^2 
( \mathring{g}^{ik} \partial_k {\cal Y}^A)  (\mathring{g}^{jl} \partial_l {\cal Y}^B)\,{\cal Y}^C {\cal Y}^D\,
\left( m^{AB}  m^{CD}  -  m^{AC} m^{BD}
\right)
\label{liftg110}
\;.
\eea
Identifying the ${\rm SO}(4)$ Killing vectors ${\cal K}_{AB}{}^i = \mathring{g}^{ij}\,\partial_j {\cal Y}_{[A} {\cal Y}_{B]}$
on the right hand side, this result reproduces the standard Kaluza-Klein ansatz for the internal metric \cite{Duff:1986hr}.
Using sphere harmonics identities
collected in appendix~\ref{sec:appS3}, we may deduce the internal metric
\bea
g_{ij} &=& \frac{\Delta^{-2}}{({\cal Y}^A{\cal Y}^B m^{AB})}\,
\partial_i{\cal Y}^C \partial_j{\cal Y}^D m_{CD}
\;,
\label{liftg110A}
\eea
together with a compact expression for the warp factor (\ref{warp0})
\bea
\Delta &=& (e^{-\varphi/2})({\cal Y}^A {\cal Y}^B m^{AB})^{-1/4} 
\;,
\label{Deltamphi}
\eea
where we recall the definition (\ref{mm20}) of the 3D scalar $\varphi$.
The latter may be used to simplify the reduction formulas (\ref{dila0})--(\ref{liftg110A}) as
\bea
e^\phi &=& \Delta^{-2}\,e^{-2\varphi} 
\;,\nonumber\\
B_{ij} &=&
- 2\,\mathring{\omega}_{ijk}
\left(
\mathring{\zeta}^k -  \mathring{g}^{kl} \partial_l \,
{\rm log}\,\Delta \right)
\;,\nonumber\\
g_{ij} &=&  \Delta^{2}\,e^{\varphi}\, 
\partial_i{\cal Y}^C \partial_j{\cal Y}^D \tilde{m}_{CD}
\;.
\label{gijN11}
\eea
We thus obtain a compact form of the full 6D metric
\bea
ds_6^2 &=&
e^{\varphi} \left(
\Delta^{-2}e^{-\varphi}\,g_{\mu\nu}(x) dx^\mu dx^\nu + \Delta^{2}\,e^{\varphi}\, 
\partial_i{\cal Y}^C \partial_j{\cal Y}^D m_{CD}\,d{y}^i  \,d{y}^j
\right)
\;,
\eea
and may also compute the internal component of the 3-form field strength
\bea
3\,\partial_{[i} B_{jk]} &=&
\mathring{\omega}_{ijk}
\left(
2\,e^{4\varphi}\Delta^8\,{\cal Y}^A m^{AC}m^{CB}{\cal Y}^B
- e^{2\varphi}\Delta^4\,m^{AA}
\right)
\;.
\eea
We may compare these results to the reduction formulas found in 
\cite{Cvetic:2000dm,Deger:2014ofa} for the ${\cal N}=(1,0)$ subsector and find precise agreement upon
applying the dictionary
\bea
{\cal Y}^A &\longleftrightarrow& \mu^i\;,\nonumber\\
m^{AB} &\longleftrightarrow& T_{ij} \;,\nonumber\\
e^{\varphi} &\longleftrightarrow& ({\rm det}\,T)^{1/4}\;,\nonumber\\
\Delta^{-2}e^{-\varphi} &\longleftrightarrow& \Delta^{1/2}\;,\nonumber\\
e^\phi &\longleftrightarrow& e^{-\sqrt{2}\,\varphi/2}
\;.
\eea
The present construction extends these formulas to the full ${\cal N}=(1,1)$ theory.
The additional matter is made from ${\cal N}=(1,0)$ vector multiplets, whose reduction
formulas are extracted from (\ref{compM11}) as
\bea
g^{-1}\,g^{ij}\,A_j{}^\alpha &=&
\mathring{g}^{-1} \mathring{g}^{ij} ( \partial_j {\cal Y}^C )\,  {\cal Y}^A {\cal Y}^B  
\left(m^{AB}   M^{C\alpha}
-  
m^{AC} M^{B\alpha}   \right)
\;,
\label{liftA0}
\eea
which upon combination with (\ref{gijN11}) and after some computation reduces to
the simple formula
\bea
A_i{}^\alpha &=& 
(\partial_i{\cal Y}^A)
\, \phi_{A}{}^{\alpha}
\;,
\label{Ai11}
\eea
showing that in particular the internal field strengths vanish
\bea
F_{ij}{}^\alpha &=& 2\,\partial_{[i} A_{j]}{}^\alpha ~=~ 0
\;.
\eea
Similarly, we extract the reduction formula for the dual 2-form $\tilde{B}_{ij}$ upon
combining (\ref{compM11}) with all previously obtained reduction formulas, and find
\bea
 \tilde B_{ij}&=&
-2\,\alpha\, \mathring{\omega}_{ijk} \mathring{\zeta}^k 
\qquad\Longrightarrow\qquad
3\,\partial_{[i} \tilde{B}_{jk]} ~=~
-2\,\alpha\, \mathring{\omega}_{ijk}
\;.
\eea
Finally, we may work out the reduction formula for the internal
components of the 6D 3-form (dual to the 6D vector fields) as
\bea
a_{ijk}{}^\alpha
&=&
-\mathring{\omega}_{ijk}\,e^{2\varphi}\,\Delta^4\,
{\cal Y}^A\,  m^{AB}\,\phi_B{}^{\alpha}
\;.
\label{aijk11}
\eea
Formulas (\ref{Ai11}) and (\ref{aijk11}) show that in the case of 3D constant scalar solutions,
all 6D vector field strengths vanish, such that the embedding of the ${\cal N}=(1,0)$ theory into the 
${\cal N}=(1,1)$ theory remains rather trivial.
This reflects the fact that the potential (\ref{pot11}) does not carry the additional 3D scalar
fields $\phi_{A\alpha}$ and thus coincides with the potential of the truncation to the quarter-maximal theory of
\cite{Cvetic:2000dm,Deger:2014ofa}. In contrast, for solutions with running scalars, such as 3D RG flows
in the potential (\ref{pot11}), these formulas describe non-trivial 6D gauge fields.

%%%%%%%%%%%%%%%%%%%%%%%%%%%%%%%

\section{${\cal N}=(2,0)$ uplift formulas}
\label{sec:N20uplift}

%%%%%%%%%%%%%%%%%%%%%%%%%%%%%%%

In this section, we repeat the analysis for the reduction of the ${\cal N}=(2,0)$ theory.
As already reflected by the richer structure of the 3D potential (\ref{pot20}),
in this case the uplift formulas to six dimensions constitute a rather non-trivial 
extension of the formulas \cite{Cvetic:2000dm,Deger:2014ofa} for the quarter-maximal truncation.

\subsection{Twist matrix}

The twist matrix describing the consistent truncation of 6D ${\cal N}=(2,0)$ supergravity
has been given in \cite{Hohm:2017wtr} in terms of the same geometrical data introduced
in section \ref{subsec:N11twist} above. 
Let us recall that the coordinates $y^i$ relevant for 6D ${\cal N}=(2,0)$ supergravity have been
identified  in (\ref{der20}) above.
In a basis where the `curved index' $M$ is decomposed according to (\ref{basis1}), 
and the `flat index' $\bar{M}$ is decomposed in the basis (\ref{GL4SO4}),
the associated twist matrix is given by
\bea
{\cal U}_M{}^{\bar M} &=&
\left(
\begin{array}{c:c:c}
{\cal U}^i{}^A & {\cal U}^i{}_A &0  \\ \hdashline 
{\cal U}_i{}^A & {\cal U}_i{}_A  &0 \\ \hdashline
{\cal U}_0{}^A &{\cal U}_{0\,A}  &0\\ \hdashline
{\cal U}_{\bar{0}}{}^A &{\cal U}_{\bar{0}A}&0 \\ \hdashline
0&0& \delta_{\alpha\beta}
\end{array}\right)
\nonumber\\
 &=&
\left(
\begin{array}{c:c:c}
\mathring{g}^{1/2}\left(
\mathring{g}^{ij}\partial_j {\cal Y}^A 
+2\,\mathring{\zeta}^i{\cal Y}^A\right)
& 
2\,\alpha\,\mathring{g}^{1/2}\left(
\mathring{\zeta}^i  {\cal Y}^A-2\,\mathring{\zeta}^i  \,\mathring{\zeta}^j\,\partial_j{\cal Y}^A  \right) &0
\\ \hdashline 
0 & \mathring{g}^{-1/2}\,\partial_i {\cal Y}^A   &0
\\ \hdashline
\frac1{\sqrt{2}} \,  {\cal Y}^A &\frac1{\sqrt{2}} \, \left( 
{\cal Y}^A-2\,(1+\alpha)\,\mathring{\zeta}^i\,\partial_i{\cal Y}^A  \right)  &0\\ \hdashline
\frac1{\sqrt{2}} \,  {\cal Y}^A &-\frac1{\sqrt{2}} \, \left(
{\cal Y}^A-2\,(1-\alpha)\,\mathring{\zeta}^i\,\partial_i{\cal Y}^A  \right)&0 \\ \hdashline
0&0& \delta_{\alpha\beta}
\end{array}\right)
\;.
\label{twist20}
\eea
Again, the free parameter $\alpha$ can (up to sign) be absorbed into a shift of the 6D dilaton.

\subsection{Uplift formulas for the 3D scalar sector}

\subsubsection{Metric}
\label{subsubsec:metric}

Combining the embedding (\ref{compM}) of the internal metric $g_{ij}$ into the scalar matrix 
with the twist ansatz (\ref{Scherk-Schwarz}) and the twist matrix (\ref{twist20}), we read off
\bea
g_{ij} &=& 
\Delta^2\,\partial_i{\cal Y}^A \,\partial_j{\cal Y}^B \,m^{AB}
\;,
\label{liftg20}
\eea
where we have defined
\bea
\Delta ~=~\Delta(x,y) &\equiv& \frac{{g}^{1/2}}{\mathring{g}^{1/2}}
~=~
e^{-\varphi} 
\left(
 {\cal Y}^A {\cal Y}^B {m}_{AB}
\right)^{-1/4}
\;.
\label{Delta20}
\eea
The matrix $m^{AB}$ denotes the ${\rm GL}(4)$ matrix constituting a $4\times4$ block
of the matrix $M_{\bar M\bar N}$ (\ref{paramM}) parametrizing the 3D coset space (\ref{coset}),
the matrix $m_{AB}$ is its inverse.
Some algebraic manipulation (c.f.\ (\ref{liftg110}) above) yields the explicit form of the inverse metric
\bea
g^{ij} &=&
e^{4\varphi} \,\Delta^2\,
( \mathring{g}^{ik} \partial_k {\cal Y}^A)  (\mathring{g}^{jl} \partial_l {\cal Y}^B)\,{\cal Y}^C {\cal Y}^D\,
\left({m}_{AB} {m}_{CD} - {m}_{AC} {m}_{BD} \right)
\;.
\label{ginv20}
\eea
Comparison to (\ref{liftg110}), (\ref{Deltamphi}), and (\ref{gijN11}) above shows precise agreement
with the reduction formulas obtained for the ${\cal N}=(1,1)$ theory upon redefinition (\ref{flipM}) of the 3D fields.

\subsubsection{2-forms}

In the same way, we extract the reduction formulas for the 6D 2-forms via the dictionary (\ref{compM}).
With the explicit form of the twist matrix (\ref{twist20}), after some computation and use of
the explicit formulas (\ref{liftg20}), (\ref{ginv20}), this gives rise to the expressions
\bea
B_{ij}{}_\alpha&=& 
\frac12\,\Delta^{-2}\,\mathring{\omega}_{ijk}\,\mathring{g}^{kl} \,
 \partial_l \left(\Delta^2 {\cal Y}^A   \phi_{A\alpha}\right)
\;,
\label{bb1}
\eea
for the ${\rm SO}(4)$ vector of 2-forms, and
\bea
\sqrt{2}\,B_{ij}{}^{0}{}
&=&
  -(1+\alpha)\,  \mathring{\omega}_{ijk}\,
 \mathring{\zeta}^k 
+
\mathring{\omega}_{ijk}\,
 \mathring{g}^{kl} \partial_l\, {\rm log}\,\Delta
+\frac{1}{8\,\Delta^4}\,\mathring{\omega}_{ijk}\,
\mathring{g}^{kl} \partial_l\left( \Delta^4 \, (\phi\phi)_{AB}\,{\cal Y}^A{\cal Y}^B \right)
\;,
\nonumber\\[2ex]
\sqrt{2}\,B_{ij}{}^{\bar0}{}
&=&
  -(1-\alpha)\,  \mathring{\omega}_{ijk}\,
 \mathring{\zeta}^k 
+
\mathring{\omega}_{ijk}\,
 \mathring{g}^{kl} \partial_l\, {\rm log}\,\Delta
-
\frac{1}{8\,\Delta^4}\,\mathring{\omega}_{ijk}\,
\mathring{g}^{kl} \partial_l\left( \Delta^4 \, (\phi\phi)_{AB}\,{\cal Y}^A{\cal Y}^B \right)
\;,\;\;
\label{bb2}
\eea
for the remaining two 2-forms.
For later use, it will be interesting to explicitly compute the associated field strengths
$H_{ijk}{}^{\bar{a}}=3\,\partial_{[i} B_{jk]}{}^{\bar{a}}$:
\bea
H_{ijk\,\alpha}&=& 
-\frac1{2}\, \mathring{\omega}_{ijk}\,\tilde\Delta^{4}
\left(
 {m}_{BB}\,{\cal Y}^A
+{m}_{AB}  \, {\cal Y}^B  
-2\,
 \tilde\Delta^{4} \,{\cal Y}^D  {m}_{DC} {m}_{CB} {\cal Y}^B {\cal Y}^A \right)
  \phi_{A\alpha}
 \;,
\nonumber\\[2ex]
\sqrt{2}\,H_{ijk}{}^{0}
&=&
- \mathring{\omega}_{ijk}
\left(\alpha   +\frac1{2}\, \tilde\Delta^{4}\, {m}_{AA}
-\tilde\Delta^{8}\,  {\cal Y}^A  {m}_{AC} {m}_{CB} 
   {\cal Y}^B 
\right)
\nonumber\\
&&{}
+\frac{1}{4}\,  \mathring{\omega}_{ijk}
\left(
 \delta^{AB}-\tilde\Delta^{4}\, ({m}_{CC} {\cal Y}^A  + 2 \,{\cal Y}^C  {m}_{AC})\, {\cal Y}^B 
\right)
(\phi\phi)_{AB}\, 
\nonumber\\
&&{}
+\frac{1}{2}\, \mathring{\omega}_{ijk}\,\tilde\Delta^{8}\,
 {\cal Y}^A{\cal Y}^B \, {\cal Y}^C  {\cal Y}^E  {m}_{CD} {m}_{DE} 
 \, (\phi\phi)_{AB}
\;,\nonumber\\[2ex]
\sqrt{2}\,H_{ijk}{}^{\bar 0}
&=&
 \mathring{\omega}_{ijk}
\left(\alpha   -\frac1{2}\, \tilde\Delta^{4}\, {m}_{AA}
+\tilde\Delta^{8}\,  {\cal Y}^A  {m}_{AC} {m}_{CB} 
   {\cal Y}^B 
\right)
\nonumber\\
&&{}
-\frac{1}{4}\, \mathring{\omega}_{ijk}
\left(
 \delta^{AB}-\tilde\Delta^{4}\, ({m}_{CC} {\cal Y}^A  + 2 \,{\cal Y}^C {m}_{AC})\, {\cal Y}^B 
\right)
(\phi\phi)_{AB}\, 
\nonumber\\
&&{}
-\frac{1}{2}\,\mathring{\omega}_{ijk}\, \tilde\Delta^{8}\,
 {\cal Y}^A{\cal Y}^B \, {\cal Y}^C  {\cal Y}^E  {m}_{CD} {m}_{DE} 
 \, (\phi\phi)_{AB}
\;,
\label{H0b}
\eea
where we have defined the rescaled $\tilde\Delta\equiv e^\varphi\Delta$\,.

\subsubsection{Scalars}

Eventually, we can compute the 6D scalar fields from the last line of (\ref{compM})
upon subtracting the $B^2$ term using explicit expressions from (\ref{bb1}), (\ref{bb2}) above.
The five 6D scalars sit in a coset space ${\rm SO}(1,5)/{\rm SO}(5)$ which we parametrize
by a symmetric positive definite matrix ${\ss M}^{\bar a\bar b}$.
Evaluation of (\ref{compM}) yields the various components of this matrix as
\bea
{\ss M}^{00} & = &  \frac{1}{8} \left( 4\,\tilde\Delta^{-4}  
+ 4\,{\cal Y}^{A} (\phi\phi)_{AB} {\cal Y}^{B} + \tilde\Delta^{4}  \left( 
2+  {\cal Y}^{A} (\phi\phi)_{AB} {\cal Y}^{B} \right)^2 \, \right)
\,, \nonumber \\
{\ss M}^{0\bar0} & = &  \frac{1}{8}  \left( 
4\,\tilde\Delta^{4} - \left(  2\, \tilde\Delta^{-2} + \tilde\Delta^{2} \,{\cal Y}^{A} (\phi\phi)_{AB} {\cal Y}^{B} \right)^2
\right)
\,, \nonumber \\
{\ss M}^{\bar0\bar0} & = &  \frac{1}{8} \left( 4\,\tilde\Delta^{-4}
+4\, {\cal Y}^{A} (\phi\phi)_{AB} {\cal Y}^{B} + \tilde\Delta^{4}  \left( 
2 -  {\cal Y}^{A} (\phi\phi)_{AB} {\cal Y}^{B} \right)^2 \, \right)
\,,
\label{mmm}
\eea
for the $2\times 2$ block in $(0, \bar0)$ directions and
\bea
{\ss M}^{0}{}_{\alpha} &=& \frac{1}{2\sqrt{2}} \left(
2+ \tilde\Delta^{4} \left(2 + {\cal Y}^{C} (\phi\phi)_{CD} {\cal Y}^{D} \right) \right)
  {\cal Y}^{A}  \phi_{A\alpha}
\,, \nonumber \\
{\ss M}^{\bar 0}{}_{\alpha} &=& \frac{1}{2\sqrt{2}} \left(
-2+ \tilde\Delta^{4} \left(2 -  {\cal Y}^{C} (\phi\phi)_{CD} {\cal Y}^{D} \right) \right)
 {\cal Y}^{A}  \phi_{A\alpha}
\,, \nonumber \\
{\ss M}_{\alpha\beta} &=& \delta_{\alpha\beta} + \tilde\Delta^{4}  \, 
{\cal Y}^{A} \, \phi_{A\alpha} \, \phi_{B\beta} \, {\cal Y}^{B} 
\,,
\label{mmm1}
\eea
for the remaining components.

\subsection{Uplift formulas for the 3D vector sector}

Building on the dictionary (\ref{vectorsN20}), we may also give the uplift of the 3D vector fields.
We recall from section \ref{subsec:3D}, (in particular (\ref{connectionAB})) that the 3D Lagrangian
carries 12 vectors fields: 6 $A_\mu{}^{AB}$ and the 6 antisymmetric combinations $A_{\mu\,A}{}^B-A_{\mu\,B}{}^A$.
Moreover, in the 3D gauge we are using (in which scalars $\phi_{AB}$ are set to zero), 
the vector fields $A_\mu{}^{A}{}_{B}$ can be eliminated by means of their algebraic field equations
in terms of scalar currents and the field strengths $\star F^{AB}$\,.

For the off-diagonal block of the 6D metric, we thus find
\bea
g^{ij} g_{j\mu} ~=~
\frac12\,\varepsilon^{ijk}\,{\cal A}_{\mu\,jk} &=&
\frac12\,\mathring{\omega}^{ijk}\,
\partial_j {\cal Y}^A \partial_k {\cal Y}^B\,
A_\mu{}^{AB}
~=~
{\cal K}_{AB}{}^i\,\tilde{A}_\mu{}^{AB}
\;,
\label{offdiag}
\eea
in terms of 
the 3D vector fields from (\ref{Atilde}) and
the ${\rm SO}(4)$ Killing vectors ${\cal K}_{AB}{}^i = \mathring{g}^{ij}\,\partial_j {\cal Y}_{[A} {\cal Y}_{B]}$,
and where we have used the relation (\ref{useEps}).
This consistently reproduces the standard Kaluza-Klein ansatz for the vector fields \cite{Duff:1986hr},
such that upon combination with the result of section~\ref{subsubsec:metric}, the full 6D metric  takes the form
\bea
ds_6^2 &=&
\Delta^{-2}\,g_{\mu\nu}(x) \, dx^\mu dx^\nu + g_{ij}\,D{y}^i  \,D{y}^j
\;,
\label{metric6Dfull}
\eea
with
\bea
D{y}^i &=& dy^i + \tilde{A}_\mu{}^{AB} {\cal K}^i{}_{AB}\,dx^\mu
\;.
\eea
Similarly, we can work out the reduction formulas for the off-diagonal blocks of the 6D 2-forms,
leading to
\bea
B_{\mu i\,0}&=&
{\cal A}_{\mu i\,0} = 
\frac{1}{\sqrt{2}}
\left(
\partial_i {\cal Y}^A\, {\cal Y}^B\left(A_\mu{}^{AB}+ A_\mu{}^{A}{}_{B}\right)
-2\,(1+\alpha)\,\mathring{\zeta}^k \partial_i {\cal Y}^A\, \partial_k {\cal Y}^B
 A_\mu{}^{AB}
\right)
\;,
\nonumber\\[2ex]
B_{\mu i\,\bar0}&=&{\cal A}_{\mu i\,\bar 0} = 
\frac{1}{\sqrt{2}}
\left(
\partial_i {\cal Y}^A\, {\cal Y}^B\left(A_\mu{}^{AB}- A_\mu{}^{A}{}_{B}\right)
-2\,(1-\alpha)\,\mathring{\zeta}^k \partial_i {\cal Y}^A\, \partial_k {\cal Y}^B
 A_\mu{}^{AB}
\right)
\;,
\nonumber\\[2ex]
B_{\mu i}{}^{\alpha}&=&
{\cal A}_{\mu i}{}^{\alpha} = 
\partial_i {\cal Y}^A\,A_\mu{}^{A\alpha}
\;.
\label{bbb}
\eea
Note that the vector fields $A_\mu{}^{A\alpha}$ do not appear in the Lagrangian (\ref{LCS}), can be
defined on-shell and subsequently be set to zero by a suitable (tensor) gauge transformation.
The complete 6D 2-forms are then given by
\bea
B_{\bar a} &=&
\frac12\,B_{ij\,\bar a} \,Dy^i \wedge Dy^j + B_{\mu i\,\bar a} \,dx^\mu \wedge Dy^i + \frac12\,B_{\mu\nu\,\bar a} \, dx^\mu \wedge dx^\nu
\;, 
\eea 
where the first two terms have been given in (\ref{bb1}), (\ref{bb2}) and (\ref{bbb}), respectively,
while the missing components $B_{\mu\nu\,\bar a}$ 
are most conveniently obtained directly from the 6D tensor self-duality equations
which allow to express their field strengths $H_{\mu\nu\rho\,\bar a}$ in terms of the associated $H_{ijk\,\bar a}$,
computed in (\ref{H0b}).

%%%%%%%%%%%%%%%%%%%%%%%%%%%%%%%

\section{Some explicit uplifts}
\label{sec:uplifts}

%%%%%%%%%%%%%%%%%%%%%%%%%%%%%%%

In order to illustrate and check the non-linear uplift formulas obtained, we
will now use them to uplift some of the AdS$_3$ solutions corresponding to the
stationary points of the 3D scalar potential to full solutions of 6D supergravity.

%%%%%%%%%%%%%%%%%%%%%%%%%%%%%%%

\subsection{6D ${\cal N}=(2,0)$ bosonic field equations}

%%%%%%%%%%%%%%%%%%%%%%%%%%%%%%%

The field equations of 6D ${\cal N}=(2,0)$ supergravity, coupled to a tensor multiplet, 
have been given e.g.\ in \cite{Romans:1986er,Riccioni:1997np}.
Apart from a metric, five selfdual and an anti-selfdual two-form,
 they feature five scalars sitting in a coset space ${\rm SO}(1,5)/{\rm SO}(5)$ which parametrize
a symmetric positive definite matrix ${\ss M}^{\bar a\bar b}$.
In our notation the field equations read
\bea
\nabla_{\hat\mu} \left( {\ss M}^{\bar a \bar c} \partial^{\hat\mu} {\ss M}_{\bar c \bar d}\right)
\eta^{\bar d \bar b} 
&=&
-\frac19\,\sqrt{G}^{-1}\,\varepsilon^{\hat\mu\hat\nu\hat\rho\hat\lambda\hat\sigma\hat\tau}\,
H_{\hat\mu\hat\nu\hat\rho}{}^{\bar a} H_{\hat\lambda\hat\sigma\hat\tau}{}^{\bar b}
\;,
\nonumber\\
R_{\hat\mu\hat\nu}-\frac12\,R^{(6)}\,g_{\hat\mu\hat\nu} &=&
-\frac18\,\partial_{\hat\mu} {\ss M}_{\bar a \bar b} \partial_{\hat\nu} {\ss M}^{\bar a \bar b}
+\frac1{16}\,g_{\hat\mu\hat\nu}\,
\partial_{\hat\rho} {\ss M}_{\bar a \bar b} \partial^{\hat\rho} {\ss M}^{\bar a \bar b}
+ 
\frac12\,H_{\hat\mu\hat\rho\hat\sigma}{}^{\bar a} H_{\hat\nu}{}^{\hat\rho\hat\sigma\,\bar b}\,{\ss M}_{\bar a\bar b}
\;,
\label{Einstein}
\eea
together with the 6D self-duality equations
\begin{equation}
\star H_{\bar a}  = {\ss M}_{\bar{a}\bar{b}} \, H^{\bar{b}}
\,.
\label{sd3f}
\end{equation}
We use indices $\hat\mu, \hat\nu, \dots = 0, \dots, 5$, to denote the curved 6D space-time indices.

%%%%%%%%%%%%%%%%%%%%%%%%%%%%%%%

\subsection{One-parameter deformation of AdS$_3\times S^3$}
\label{subsec:lift1}

%%%%%%%%%%%%%%%%%%%%%%%%%%%%%%%

As a first example, we give the 6D uplift of the non-supersymmetric but stable
one-parameter family of AdS$_3$ solutions (\ref{family})
located at
\bea
m^{AB} &=&{\rm diag}\left\{e^\eta,e^\eta,e^{-\eta},e^{-\eta}\right\}\;,\quad
\phi_{A\alpha} ~=~0
\;,
\eea
into 6D ${\cal N}=(2,0)$ supergravity.
With an explicit parametrization of the $S^3$ sphere harmonics as
\bea
&&{\cal Y}^A = \{ u^\alpha {\rm cos}\,\theta, v^\alpha {\rm sin}\,\theta \} 
\;,\qquad
u^\alpha u^\alpha = 1 = v^\alpha v^\alpha
\;,\nonumber\\
&&{}
u^\alpha = ({\rm cos}\, \xi_1, {\rm sin}\, \xi_1)\;,\quad
v^\alpha = ({\rm cos}\, \xi_2, {\rm sin}\, \xi_2)
\;,
\label{param}
\eea
the warp factor $\Delta$ is given by (\ref{Delta20})
\bea
\Delta ~=~
\left(
{\rm cosh}\,\eta-{\rm cos}(2\theta)\,{\rm sinh}\,\eta  \right)^{-1/4}
\;.
\eea
The six-dimensional metric is then obtained from (\ref{metric6Dfull}) as
a warped product of AdS$_3$ and a deformed sphere $S^3$
\bea
ds^2_6&=&
\Delta^{-2}\left(ds^2_{{\rm AdS}_3}
+ d\theta^2 \right)
+
e^{\eta}\, \Delta^2 \,{\rm cos}^2\theta\,d\xi_1^2
+e^{-\eta}\, \Delta^2 \,{\rm sin}^2\theta  \,d\xi_2^2
\;,
\eea
with the two surviving U(1) isometries corresponding to rotations along $\xi_1$ and $\xi_2$.
The full 6D curvature scalar follows as
\bea
R^{(6)} &=& \Delta^{10}\,{\rm sin}^2(2\theta)\,{\rm sinh}^2\eta
\;.
\eea
The SO(1,5) scalars are computed from (\ref{mmm}) as
\bea
{\ss M}^{00} & = &  \frac{1}{2} \left( \Delta^{4} 
 + \Delta^{-4} \right) ~=~{\ss M}^{\bar0\bar0} 
\,,\quad
{\ss M}^{0\bar0} ~ = ~  
\frac{1}{2}  \left( \Delta^{4}  - \Delta^{-4}  \right) 
\,,
\nonumber\\
{\ss M}_{\alpha\beta} &=& \delta_{\alpha\beta} 
\;,\quad
{\ss M}^{0}{}_{\alpha} ~=~ 0~=~
{\ss M}^{\bar 0}{}_{\alpha} 
\,,
\eea
and the components of the 3-form field strengths along the $S^3$ directions 
follow from (\ref{H0b}) to be
\bea
\sqrt{2}\,H_{ijk}{}^{0}
=
- \mathring{\omega}_{ijk}
\left( \Delta^{8} +\alpha  
\right)
\;,\quad
\sqrt{2}\,H_{ijk}{}^{\bar 0}
=
- \mathring{\omega}_{ijk}
\left( \Delta^{8} -\alpha  
\right)\;,
\quad
H_{ijk\,\alpha}= 0
 \;.
\eea
The remaining components of the 6D field strengths can then be determined
by imposing the 6D self-duality equations (\ref{sd3f}), giving rise to
\bea
H^{0} &=& - \frac1{\sqrt{2}}
\left(
\left( \Delta^{8} +\alpha  
\right) \mathring\omega_{S^3}+
(\alpha+1)\,\mathring\omega_{\rm AdS}
\right)
\;,
\nonumber \\
H^{\bar0} &=& 
- \frac1{\sqrt{2}}
\left(
\left( \Delta^{8} -\alpha  
\right) \mathring\omega_{S^3}+
(\alpha-1)\,  \mathring\omega_{\rm AdS}
\right)
\;,
\eea
and vanishing $H_{\alpha}$. The field strengths are given
in terms of the volume forms $\mathring\omega_{S^3}$, $\mathring\omega_{\rm AdS}$
of unit-length $S^3$ and AdS$_3$, respectively.
The Bianchi identities constitute a non-trivial consistency check of this result.
Furthermore, it is straightforward to check that all 6D second order field equations (\ref{Einstein})
are indeed satisfied for $\alpha^2=1$.

%%%%%%%%%%%%%%%%%%%%%%%%%%%%%%%

\subsection{Uplift of an AdS$_3$ vacuum}

%%%%%%%%%%%%%%%%%%%%%%%%%%%%%%%

As a second example, let us work out the 6D uplift of the stationary point $(i)$
(\ref{stationary}) of the potential (\ref{pot20}). 
Although this solution is unstable as an AdS$_3$ vacuum, thus not of immediate interest,
the fact that its uplift solves all 6D field equations constitutes a non-trivial 
consistency check to our uplift formulas.
Recall that the location of this solution is specified by (\ref{soli}) with $\mathfrak{m} \equiv 3/2$.
Using the explicit parametrization introduced earlier (\ref{param}) for the sphere harmonics, one
now finds a constant warp factor \( \Delta = \mathfrak{m}^{-3/4} \). 
Then the six-dimensional metric is readily obtained 
as
\bea
ds_6^2 &=&
\mathfrak{m}^{3/2} \,d\mathring{s}^2_{{\rm AdS}}
+ \mathfrak{m}^{-1/2} \,d\mathring{s}^2_{S^3}\;.
\eea
The Ricci tensor of this metric can be conveniently given as
\bea
R_{\hat\mu\hat\nu} \,dx^{\hat\mu} dx^{\hat\nu} &=&
-\frac{2}{\ell^2}\,d\mathring{s}^2_{{\rm AdS}}
+2\,d\mathring{s}^2_{S^3}
\;, \quad
R^{(6)} =
6\, \left( \mathfrak{m}^{1/2} - \frac{1}{\ell^2} \,\mathfrak{m}^{-3/2} \right)
\;,
\eea
which leads to
\bea
\left( R_{\hat\mu\hat\nu} -\frac12\,R^{(6)} g_{\hat\mu\hat\nu}  \right)  dx^{\hat\mu} dx^{\hat\nu} &=&
3\,\mathfrak{m}^{2} \left( \frac{1}{3\,\mathfrak{m}^{2} \, \ell^2} - 1 \right) d\mathring{s}^2_{{\rm AdS}}
+
 \left( \frac{3}{\mathfrak{m}^{2} \, \ell^2} - 1 \right) d\mathring{s}^2_{S^3}
 \;,
\eea
for the Einstein tensor.
The scalars are obtained from (\ref{mmm}) and (\ref{mmm1}) as
\bea
&&{\ss M}^{00} =  \frac{1}{2} \left( \frac{9}{4 \, \mathfrak{m}} + 1 + \mathfrak{m} \right) = 2
\,,\;\;
{\ss M}^{0\bar0} = \frac{1}{2} \left( \mathfrak{m} - \frac{9}{4 \, \mathfrak{m}}  \right) = 0
\,,\;\;
{\ss M}^{\bar0\bar0} = \frac{1}{2} \left( \frac{9}{4 \, \mathfrak{m}} - 1 + \mathfrak{m} \right) = 1
\,, \nonumber \\
&&
{\ss M}^{0}{}_{\alpha} = 
\frac{1}{\sqrt{2 \,\mathfrak{m}} } \left( \mathfrak{m} + \frac{3}{2}  \right) {\cal Y}^{\alpha} = \sqrt{3} \, {\cal Y}^{\alpha} 
\,,\;\;
{\ss M}^{\bar 0}{}_{\alpha} = 
\frac{1}{\sqrt{2 \,\mathfrak{m}} } \left( \mathfrak{m} - \frac{3}{2}  \right) {\cal Y}^{\alpha} = 0
\,,\;\;
\nonumber\\
&&
{\ss M}_{\alpha\beta} = \delta_{\alpha\beta} + {\cal Y}^{\alpha} {\cal Y}^{\beta} \,,
\nonumber
\eea
whereas the 3-form field strengths along the $S^3$ directions computed from (\ref{H0b})
are
\bea
H_{ijk}{}^{0} = - \sqrt{2}\,\mathring{\omega}_{ijk}
\;,\quad
H_{ijk}{}^{\bar 0} = 0
\;,\quad
H_{ijk\,\alpha} = - \frac{3}{2 \sqrt{\mathfrak{m}}} \,
{\cal Y}^{\alpha} \, \,\mathring{\omega}_{ijk} = -\frac{\sqrt{3}}{\sqrt{2}} \,
{\cal Y}^{\alpha} \, \,\mathring{\omega}_{ijk} 
\;.
\eea
The 6D self-duality equations (\ref{sd3f}) can be used to determine the full
6D field strengths as
\bea
H^{0} = - \sqrt{2} \,
\left(
\mathring\omega_{S^3}+
\frac{\mathfrak{m}^{3}}{2}\,\mathring\omega_{\rm AdS}
\right)
\;,\quad
H^{\bar 0} = 0
\;,\quad
H_{\alpha} = -\frac{\sqrt{3}}{\sqrt{2}} \,
{\cal Y}^{\alpha} \, \mathring\omega_{S^3}  \,.
\label{3fstarweps}
\eea
The determination of the 3-form field strengths by self-duality requirement renders the
Bianchi identities \( d H^{\bar a} = 0 \) non-trivial, and one verifies straightforwardly that they are satisfied. For this, it 
is crucial that \( H_{\hat\mu\hat\nu\hat\rho \, \alpha} \) has no components along the AdS$_3$ 
directions which is indeed the case.

Moreover, the different contributions to the energy-momentum tensor are given by
\bea
\left(\partial_{\hat\mu} {\ss M}_{\bar a \bar b} \partial_{\hat\nu} {\ss M}^{\bar a \bar b}
-\frac1{2}\,g_{\hat\mu\hat\nu} \,
\partial_{\hat\rho} {\ss M}_{\bar a \bar b} \partial^{\hat\rho} {\ss M}^{\bar a \bar b}
\right)  dx^{\hat\mu} dx^{\hat\nu}
&=&   6\,\mathfrak{m}^2\, d\mathring{s}^2_{{\rm AdS}}
+
2\,d\mathring{s}^2_{S^3}
\;,
\nonumber\\
\left( H_{\hat\mu\hat\rho\hat\sigma}{}^{\bar a} 
H_{\hat\nu}{}^{\hat\rho\hat\sigma\,\bar b}\,{\ss M}_{\bar a\bar b}\right)
dx^{\hat\mu} dx^{\hat\nu}
&=&
 -3\,\mathfrak{m}^2\, d\mathring{s}^2_{{\rm AdS}}
+
3\,d\mathring{s}^2_{S^3}
 \;.
\eea
From this it follows immediately that the Einstein equations (\ref{Einstein}) are verified with 
\( \mathfrak{m}^2 \ell^2 = 4/3 \).

%%%%%%%%%%%%%%%%%%%%%%%%%%%%%%%

\section{Conclusions}
\label{sec:conclusions}

In this paper we have used the framework of exceptional field theory to work out the 
consistent truncations of 6D ${\cal N}=(1,1)$ and ${\cal N}=(2,0)$ supergravity theories
on AdS$_3\times S^3$\,. The resulting three-dimensional theories are ${\rm SO}(4)$ 
gauged supergravities coupled to 4 half-maximal scalar multiplets, describing the 32 
bosonic degrees of freedom.
Employing the Scherk-Schwarz twist matrices from \cite{Hohm:2017wtr} and establishing the explicit
dictionary between ExFT fields and the 6D supergravity fields, it is straightforward to derive the 
non-linear Kaluza-Klein reduction Ans\"atze for the various 6D fields.
In the truncation to the common ${\cal N}=(1,0)$ sector, the formulas consistently reduce
to the reduction formulas from \cite{Cvetic:2000dm,Deger:2014ofa}.
The results nicely illustrate the power of the ExFT framework as
a tool in the study of consistent truncations.

The three-dimensional scalar potentials allow for a number of stationary points, most of which, 
however, turn out to be unstable by the existence of scalar directions with negative mass 
squares below the Breitenlohner-Freedman bound. Interestingly, they admit a one-parameter 
family of non-supersymmetric but stable AdS$_3$ solutions. We have given the explicit uplift 
of this family to six dimensions. Further direct applications of our uplift formulas may include 
three-dimensional solutions with non-constant scalars such as holographic RG flows in the 
scalar potentials. On a more general note, the proof of the consistent truncation to particular 
three-dimensional gauged supergravities allows to consistently restrict holographic supergravity
calculations such as \cite{Mihailescu:1999cj,Arutyunov:2000by,Giusto:2018ovt,Rastelli:2019gtj}
to a closed subsector of fields.

An immediate generalization of the results reported here is their extension to six-dimensional
supergravities with additional tensor multiplet couplings which generically arise from reductions from
ten dimensions. In the ExFT context this corresponds to an embedding ${\rm SO}(8,4)\hookrightarrow {\rm SO}(8,4+n)$
of the exceptional field theories and the associated twist matrices. Upon working out the extended dictionary
between ExFT and supergravity fields, the corresponding uplift formulas can be extracted in analogy to
the results of this paper.

It would also be highly interesting to examine if similar techniques could be employed to construct
consistent truncations involving higher massive Kaluza-Klein multiplets and leading to three-dimensional
theories of the type constructed in~\cite{Nicolai:2003ux}. This might require an extension of the present
framework to more general embeddings in the spirit of \cite{Malek:2017njj,Cassani:2019vcl}.

Finally, it would be interesting to explore to which extent similar structures can be unveiled in the context
of AdS$_3\times S^2$ truncations of the five-dimensional supergravities
obtained from compactification of M-theory on Calabi-Yau three-manifolds.

%%%%%%%%%%%%%%%%%%%%%%%%%%%%%%%

\subsubsection*{Acknowledgments}
\"{O}S is partially supported by the Scientific and Technological Research Council
of Turkey (T\"{u}bitak) Grant No. 116F137. \"{O}S would like to thank the ENS de Lyon
for hospitality and the French government for support through the SSHN scholarship
at the early stages of this work.
%%%%%%%%%%%%%%%%%%%%%%%%%%%%%%%

\section*{Appendix}

%%%%%%%%%%%%%%%%%%%%%%%%%%%%%%%

\begin{appendix}

\section{$S^3$ harmonics and identities}
\label{sec:appS3}
Here we list some of the useful identities that we have used throughout the text. 
Consider a parametrization of the unit radius $S^3$ by some coordinates ${\cal Y}^A$ 
(with $A=1, \dots, 4$) as
\bea
{\cal Y}^A {\cal Y}^A &=&1\;.
\eea
The isometries of $S^3$ can be described in terms of the ${\rm SO}(4)$ Killing vectors
\bea
{\cal K}_{AB\,i} &=& \partial_i {\cal Y}_{[A} {\cal Y}_{B]}
\;.
\eea
Then the metric of the round $S^3$ can be written in the ${\rm SO}(4)$-covariant form as
\bea
\mathring{g}_{ij} &=& 2\,{\cal K}_{AB\,i} {\cal K}_{AB\,j}
\;. 
\eea
Using these and the inverse metric $\mathring{g}^{ij}$ of the round  $S^3$, we find that
\bea
\mathring{g}^{ij} \partial_i {\cal Y}^A \partial_j {\cal Y}^B &=&
\delta^{AB} - {\cal Y}^A {\cal Y}^B
\;,
\label{dYdY}
\eea
which has proven to be of great value in the simplification of the uplift formulas throughout.
The following was also of use for the derivation of (\ref{offdiag})
\bea
\mathring{\omega}^{kij}\,
\partial_i {\cal Y}^A \partial_j {\cal Y}^B
&=&
\varepsilon_{ABCD}\,\mathring{g}^{kl}\,\partial_l {\cal Y}^C  {\cal Y}^D
\;.
\label{useEps}
\eea

\end{appendix}

%%%%%%%%%%%%%%%%%%%%%%%%%%%%%%%%%%%%%%%%%%%%%%%%%%%%

%\bibliographystyle{utphys}
%\bibliography{refs}

\providecommand{\href}[2]{#2}\begingroup\raggedright\endgroup

%%%%%%%%%%%%%%%%%%%%%%%%%%%%%%%%%%%%%%%%%%%%%%%%%%%%

\end{document}